\documentclass[%
reprint,https://www.overleaf.com/project/63fa3e58d38edcc4f777cb6a,
superscriptaddress,
 amsmath,amssymb,
 aps, prl,
]{revtex4-2}
\usepackage{hyperref}
\hypersetup{
    colorlinks=true,
    linkcolor=blue,
    citecolor=blue,
    filecolor=magenta,      
    urlcolor=blue,
    }

\usepackage{graphicx}
\usepackage{dcolumn}
\usepackage{bm}
\usepackage{graphicx, color}

\newcommand{\HL}[1]{{\color{blue} {#1}}}

\begin{document}

\preprint{APS/123-QED}

\title{Differentiating Three-Dimensional Molecular Structures using\\ Laser-induced Coulomb Explosion Imaging}

\author{Huynh Van Sa Lam}
\email{\textcolor{black}{huynhlam@phys.ksu.edu}}
\affiliation{James R. Macdonald Laboratory, Kansas State University, Manhattan, KS 66506, USA}
\author{Anbu Selvam Venkatachalam}
\affiliation{James R. Macdonald Laboratory, Kansas State University, Manhattan, KS 66506, USA}
\author{Surjendu Bhattacharyya}
\affiliation{James R. Macdonald Laboratory, Kansas State University, Manhattan, KS 66506, USA}
\author{Keyu Chen}
\affiliation{James R. Macdonald Laboratory, Kansas State University, Manhattan, KS 66506, USA}
\author{Kurtis Borne}
\affiliation{James R. Macdonald Laboratory, Kansas State University, Manhattan, KS 66506, USA}
\author{Enliang Wang}
\affiliation{James R. Macdonald Laboratory, Kansas State University, Manhattan, KS 66506, USA}
\author{Rebecca Boll}
\affiliation{European XFEL, 22869 Schenefeld, Germany}
\author{Till Jahnke}
\affiliation{European XFEL, 22869 Schenefeld, Germany}
\author{Vinod Kumarappan}
\affiliation{James R. Macdonald Laboratory, Kansas State University, Manhattan, KS 66506, USA}
\author{Artem Rudenko}
\affiliation{James R. Macdonald Laboratory, Kansas State University, Manhattan, KS 66506, USA}
\author{Daniel Rolles}
\email{\textcolor{black}{rolles@phys.ksu.edu}}
\affiliation{James R. Macdonald Laboratory, Kansas State University, Manhattan, KS 66506, USA}


\date{\today}

\begin{abstract}
Coulomb explosion imaging (CEI) with x-ray free electron lasers has recently been shown to be a powerful method for obtaining detailed structural information of gas-phase planar ring molecules [R.~Boll \textit{et al.} Nat.~Phys.~\textbf{18}, 423–428 (2022)]. In this Letter, we investigate the potential of CEI driven by a tabletop laser and extend this approach to differentiating three-dimensional (3D) structures. We study the static CEI patterns of planar and nonplanar organic molecules that resemble the structures of typical products formed in ring-opening reactions. Our results reveal that each molecule exhibits a well-localized and distinctive pattern in 3D fragment-ion momentum space. We find that these patterns yield direct information about the molecular structures and can be qualitatively reproduced using a classical Coulomb explosion simulation. Our findings suggest that laser-induced CEI can serve as a robust method for differentiating molecular structures of organic ring and chain molecules. As such, it holds great promise as a method for following ultrafast structural changes, e.g., during ring-opening reactions, by tracking the motion of individual atoms in pump-probe experiments.

\end{abstract}

\maketitle

Coulomb explosion imaging \cite{Vager1989} (CEI) is a method for imaging molecular structure by rapidly removing many electrons from the molecule, causing the highly charged parent ion to explode into multiple charged fragments due to the Coulomb repulsion between positive charges. By measuring the momenta of these fragment ions, detailed information about the molecular structure can be retrieved \cite{Kwon1996, Stapelfeldt1998}. CEI has various applications in physics and chemistry (for instance, see \cite{Pitzer2013, Herwig2013, Slater2014, Burt2018, Corrales2019, Endo2020b}), enabling new insights into chemical reactions and molecular dynamics as reviewed in \cite{Whitaker2003, Cornaggia2009, Karimi2016, Yatsuhashi2018, Hishikawa2020, Schmidt2021, Jahnke2021, Schouder2022, Li2022b, Crane2023}.

\begin{figure}
  \includegraphics[width=\columnwidth]{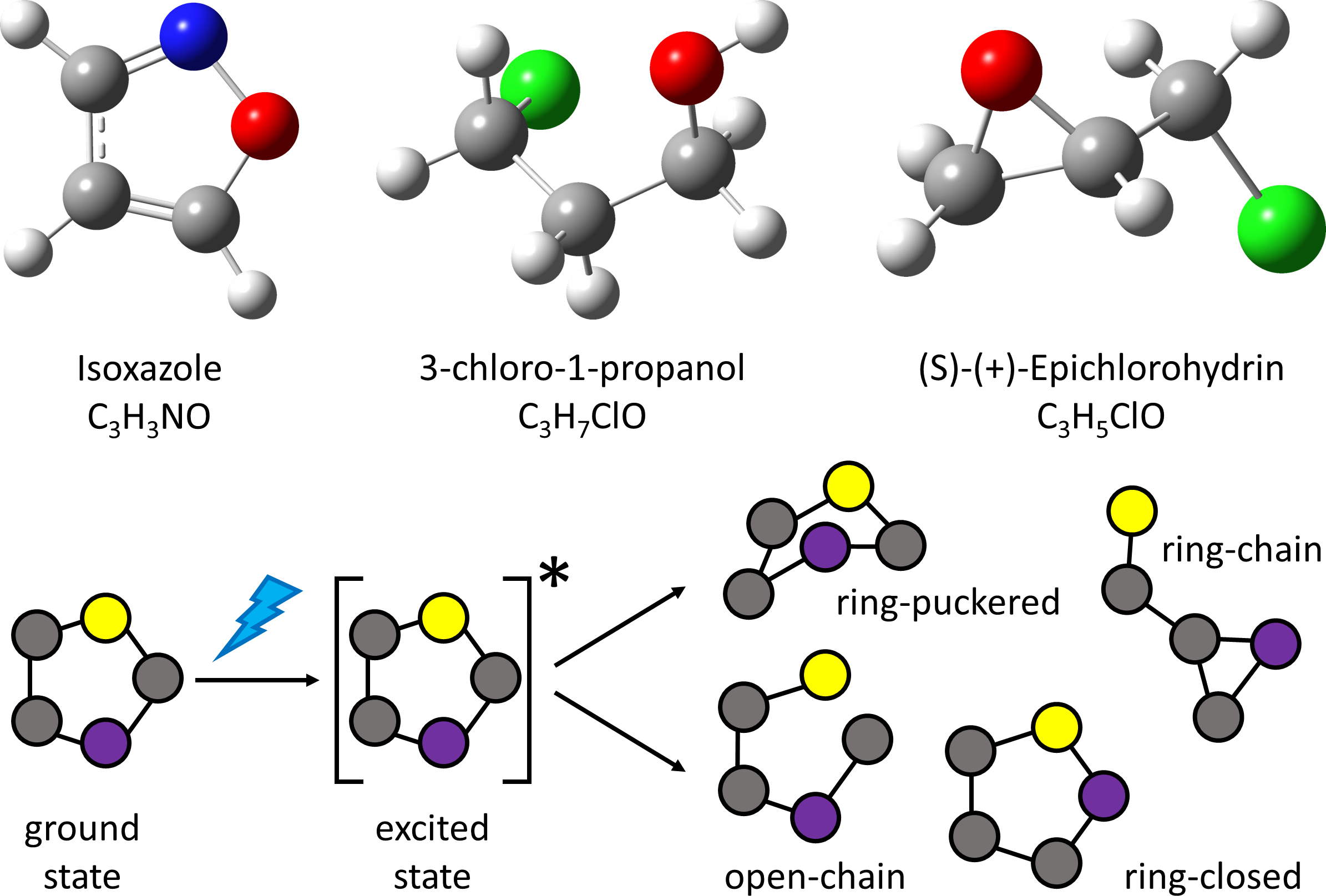}
  \caption{\textbf{Top row:} Ball-and-stick models of the three molecules studied in this work: isoxazole, 3-chloro-1-propanol, and epichlorohydrin. \textbf{Bottom row:} Schematic of a light-induced ring-opening reaction (modeled after the UV-induced ring opening of oxazole \cite{Cao2015, Cao2016}). Isoxazole, 3-chloro-1-propanol, and epichlorohydrin mimic the structures of the ring-closed, open-chain, and ring-chain products, respectively.}
  \label{fig:motivation}
\end{figure}

\begin{figure*}
  \includegraphics[width=\textwidth]{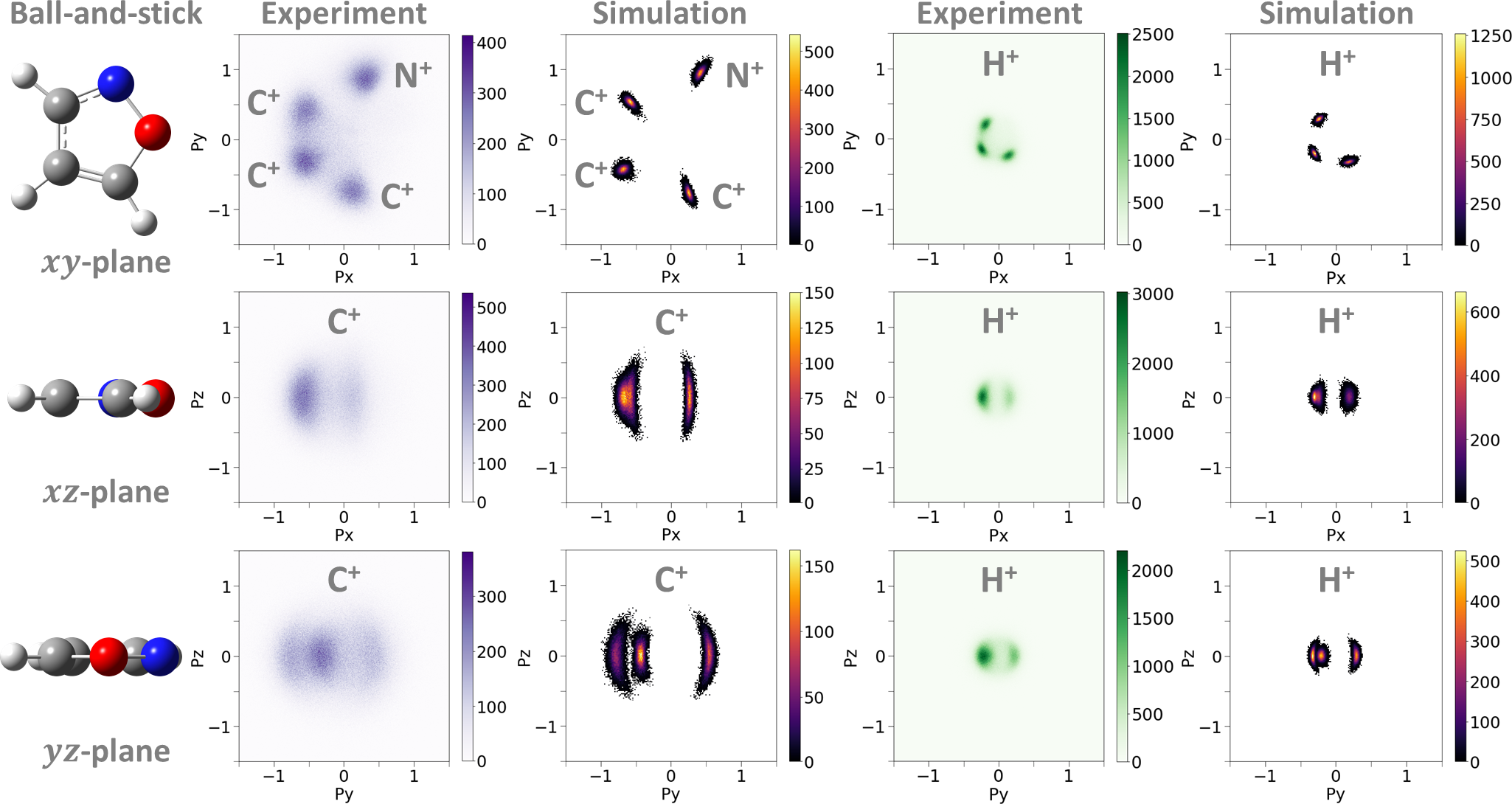}
  \caption{Measured and simulated CEI patterns of isoxazole. The first column shows the ball-and-stick model views of isoxazole in three principal planes. The plots in the second and fourth columns are projections of the measured fragment ion momenta (so-called Newton plots) from the ($\mathrm{H^+,C^+,N^+,O^+}$) 4-fold coincidence channel onto these three principal planes. The first, second, and third rows are the $xy-$plane, $xz-$plane, and $yz-$plane, respectively. For each event plotted here, the coordinate frame is rotated such that the $\mathrm{O^+}$ momentum points along the $x-$axis ($p_{O_y} = p_{O_z} = 0$) and the $\mathrm{N^+}$ momentum lies in the upper $xy-$plane ($p_{N_y} \ge 0, p_{N_z} = 0$). The momenta of $\mathrm{C^+}$ and $\mathrm{H^+}$ are plotted in this coordinate frame. No background was subtracted. All momenta are normalized to the $\mathrm{O^+}$ momentum ($|p_O| = 1$, not shown). Newton plots of $\mathrm{H^+}$ and $\mathrm{C^+}$ show localized spots corresponding to three carbons and three hydrogens of isoxazole. For laser shots where more than one carbon or hydrogen ion was detected, multiple 4-fold coincidence events were created in the analysis by making all possible combinations of the detected ($\mathrm{H^+,C^+,N^+,O^+}$) fragments. For the data visualization, the $\mathrm{C^+}$ ion counts are multiplied by three since only one out of three carbon ions is plotted for each nitrogen and oxygen ion that is plotted. The third and fifth columns show the results of classical Coulomb explosion simulations for the neutral molecule in its equilibrium geometry.}
  \label{fig:isoxazole}
\end{figure*}

\begin{figure*}
  \includegraphics[width=\textwidth]{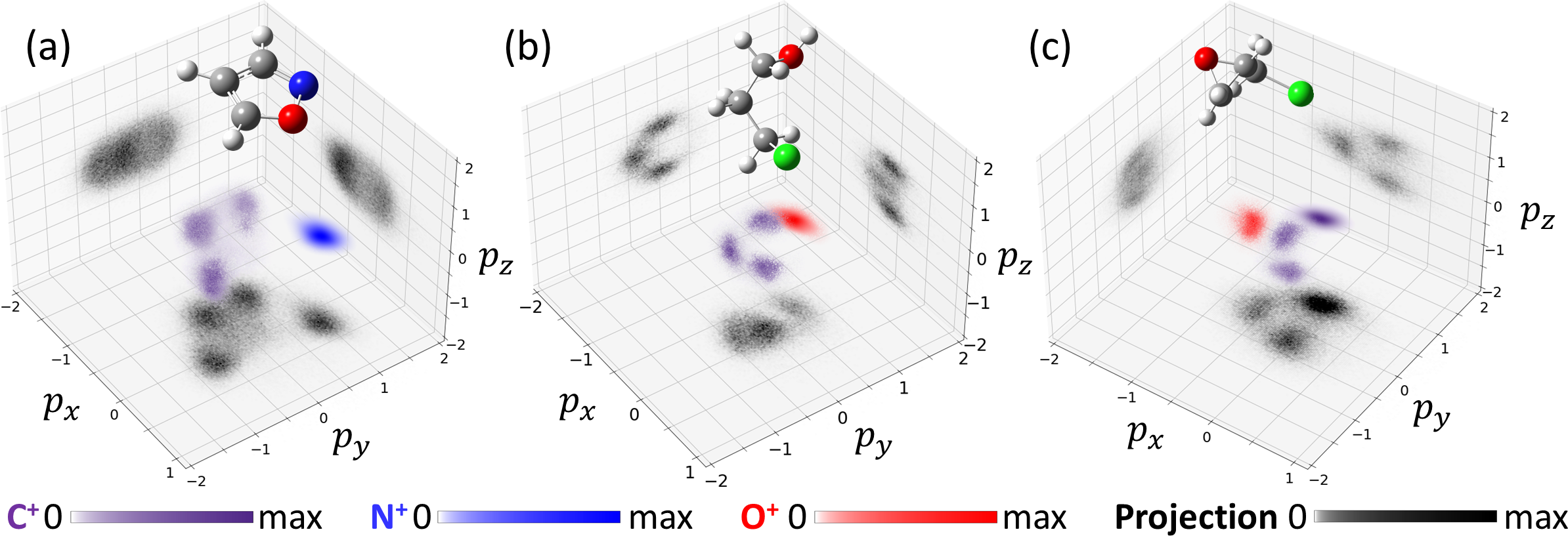}
  \caption{3D scatter plots showing the normalized measured momenta of individual ions from (a) isoxazole, (b) 3-chloro-1-propanol, and (c) epichlorohydrin from the (a) ($\mathrm{C^+,C^+,N^+,O^+}$), (b) ($\mathrm{C^+,C^+,O^+,Cl^+}$), and (c) ($\mathrm{C^+,C^+,O^+,Cl^+}$) 4-fold coincidence channels. $\mathrm{C^+}$, $\mathrm{N^+}$ and $\mathrm{O^+}$ ions in 3D scatter plots are shown in purple, blue and red, respectively, to distinguish them from projections shown in gray scale. The position of each data point is determined by its normalized momentum, while its color corresponds to the density. In (a), the recoil frame is rotated such that the momentum of $\mathrm{O^+}$ (so-called $x$-reference ion) is set as the unit vector along the $x$ axis, and the momentum of $\mathrm{N^+}$ (referred to as $xy$-reference ion) is in the upper $xy$ plane. The momenta of other ions are plotted in this coordinate frame. The $x$- and $xy$-reference ions are $\mathrm{Cl^+}$ and $\mathrm{O^+}$ in (b), and $\mathrm{Cl^+}$ and the first-detected $\mathrm{C^+}$ in (c). The $x$-reference ion is not plotted in any panel. We also do not plot the $xy$-reference ion in the $p_zp_x$ and $p_zp_y$ projections since they are simply intense lines along $p_z=0$.}
  \label{fig:3d}
\end{figure*}

Compared to other methods for imaging time-resolved structural dynamics (such as electron and x-ray diffraction), CEI provides excellent temporal resolution, high sensitivity to light atoms, and direct access to three-dimensional (3D) information. However, due to the low count rate of multiparticle coincidence measurements, most time-resolved CEI studies have focused on two- and three-body fragmentation \cite{Whitaker2003, Cornaggia2009, Karimi2016, Yatsuhashi2018, Hishikawa2020, Schmidt2021, Jahnke2021, Li2022b, Crane2023}. As a result, while CEI has been highly successful in imaging small molecules \cite{Stapelfeldt1995, Alnaser2004, Legare2005, Ergler2006, Takanashi2017, Severt2022, Howard2023}, only limited structural information has been obtained using CEI for larger molecules.

Recently, advances in high-repetition-rate intense x-ray free electron laser (XFEL) light sources, detector performance \cite{Fehre2018}, and data analysis methods have enabled the retrieval of detailed structures of gas-phase planar ring molecules with eleven atoms using CEI \cite{Boll2022}. This suggests that time-resolved CEI could be used to directly image structural changes of molecules during photochemical reactions---such as ring-opening processes---with atomic resolution. To expand the potential impact and applications, it is worth investigating the possible extension of this method to nonplanar molecules, 3D momentum space, and intense tabletop lasers, which are more widely accessible. 

In this Letter, we show that it is indeed possible to use intense near-infrared laser pulses to obtain Coulomb explosion images of molecules with 8 to 12 atoms with similar quality as those shown by Boll \textit{et al.} \cite{Boll2022}, and that this method also provides 3D structural information for nonplanar molecules. Previously, both of these aspects were only demonstrated for the case of a molecule with five different atomic constituents, where all five fragment ions were detected in coincidence \cite{Pitzer2013}. The present work, therefore, constitutes an important step towards making CEI a more generally applicable technique for probing chemical transformations involving 3D motion of atoms in carbon ring and carbon chain molecules.

As illustrated in Fig.~\ref{fig:motivation} (bottom), a prototypical ring-opening reaction can produce various final products, including vibrationally-hot ring-closed molecules and several ring-opened products. Ring-opened products are typically open-chain and ring-chain structures, with the latter containing a smaller, highly-strained ring. These products are often non-planar and have low symmetry. For example, upon absorbing one 266-nm photon, the heterocyclic molecule thiophenone can form open-chain (ketene) or ring-chain (episulfide) photoproducts or return to the ring-closed geometry \cite{Pathak2020}. Similar photoproducts are also predicted for other molecules, such as furan \cite{Fuji2010, Hua2016} and oxazole \cite{Cao2015, Cao2016}. Distinguishing these products in a time-resolved measurement remains challenging. As a first step towards a time-resolved CEI experiment targeting that goal, we investigate the static CEI patterns of isoxazole, 3-chloro-1-propanol, and epichlorohydrin molecules, which resemble the structures of the ring-closed, open-chain, and ring-chain products, respectively. The ability to distinguish a ring-puckered structure from a planar structure using CEI strongly depends on how dissimilar the two structures are in the specific molecule of interest \cite{Wang2023}, and we therefore concentrate on the other photoproduct geometries here. We show that distinctive 3D momentum patterns of these molecules can be constructed from kinematically incomplete coincidence channels (where not all the molecular constituents are detected) with only four fragment ions detected in coincidence. The experimental data are compared with CEI patterns produced from classical Coulomb explosion simulations and contrasted with the real-space geometries of the molecules. These static CEI patterns can be used as reference images to help identify the photoproducts in future time-resolved CEI experiments.

The experiment used a Ti-sapphire laser running at 3-kHz, producing near-infrared pulses of 810-nm central wavelength and 25-fs pulse duration. The pulses were focused to a peak intensity of approximately $\mathrm{10^{15}~W/cm^2}$. The molecular sample was delivered as a dilute, cold molecular beam produced by supersonic expansion with helium carrier gas. The experiment employed a double-sided velocity map imaging apparatus with time- and position-sensitive detectors for multihit coincidence imaging \cite{Robatjazi2021D, Pathak2021D}. Only the ion side of the spectrometer was used here. The fragment ions were detected in coincidence, and their momenta were determined from the measured time-of-flight (ToF) and impact position of each ion. Further details are given in Sec.~I of the Supplemental Material (SM).

Figure~\ref{fig:isoxazole} shows the ion momentum images (Newton plots) resulting from the fragmentation of the ring molecule, isoxazole, constructed from the coincidence events where $\mathrm{O^+}$, $\mathrm{N^+}$, and at least one $\mathrm{C^+}$ and one $\mathrm{H^+}$ (each identified by their different mass-to-charge ratios, see Sec.~VII of the SM) are detected. The frame of reference is defined by the $\mathrm{O^+}$ and $\mathrm{N^+}$ momenta (see caption for details), and the $\mathrm{C^+}$ and $\mathrm{H^+}$ are plotted in this frame. The maxima in the momentum distributions of the carbon, nitrogen, and hydrogen ions are well-localized and separated. Furthermore, when comparing the momentum images with the ball-and-stick model of the molecule shown on the left, the correspondence between the equilibrium molecular geometry and momentum distribution is very clear.

Note that these all-atom momentum-space patterns in isoxazole are obtained from a 4-fold coincidence channel. Although we only require one $\mathrm{C^+}$ and one $\mathrm{H^+}$ to be detected in each coincidence event, all $\mathrm{C^+}$ and $\mathrm{H^+}$ ions are visible in these momentum images containing multiple coincidence events since all the carbon ions (or all protons) are detected with similar probability. Coincidence channels where more carbon ions are detected produce similar images (see Sec.~\HL{III} of the SM). Our result shows that CEI using intense tabletop laser can provide results of clarity comparable to those obtained with an XFEL \cite{Boll2022}. In both cases, kinematically incomplete channels (e.g., detection of only 4 atomic ions from an 8-atom molecule) are sufficient to image complex molecules, indicating that in both cases, the charge-up preceding the Coulomb explosion occurs in a rapid and well-defined manner.

The third and fifth columns of Fig.~\ref{fig:isoxazole} show results of classical Coulomb explosion simulations for the equilibrium geometry. We first use the Gaussian software package \cite{g09} to optimize the geometry of the neutral molecule in its electronic ground state and calculate vibrational modes and frequencies (see SM, Sec.~VIII for the optimized geometries). After that, we use the NewtonX software \cite{Barbatti2014, NTX} to sample the initial distribution of geometries in the ground state. Finally, we perform classical Coulomb explosion simulations (CES) for the sampled geometries assuming purely Coulombic potenital initiated by instantaneously putting one point charge at each atom. Although this model overestimates the magnitudes of fragment momenta, it captures the correlations between the fragment momenta rather accurately (also see Sec.~II of the SM).

For planar molecules, the projection onto the molecular plane (e.g., the $xy$-plane for isoxazole in Fig.~\ref{fig:isoxazole}) is sufficient to give a direct visualization of the complete molecule. However, our experimental and theoretical analysis of the other two planes ($xz$ and $yz$) reveals that the Coulomb explosion of isoxazole exhibits a rather broad out-of-plane component, which reflects the distribution of geometries in the initial state of the target mapped by the Coulomb explosion process. Our data and simulations also capture the appearance of the middle carbon in the $yz$-plane (as compared to the $xz$-plane) with a smaller out-of-plane momentum spread than the other two carbons, which is a consequence of its correlated motion with the reference ions (see SM, Sec.~III for details). A 3D analysis, which considers projections on all three principal planes as shown in Fig.~\ref{fig:isoxazole}, is critical for nonplanar molecules, where each projection only provides partial information about the structure. To this end, it is natural to represent the 3D momenta as 3D scatter plots in Fig.~\ref{fig:3d} since it is easier to visualize the out-of-plane components \cite{Pitzer2013, Li2022, Bhattacharyya2022}.

\begin{figure}
  \includegraphics[width=\columnwidth]{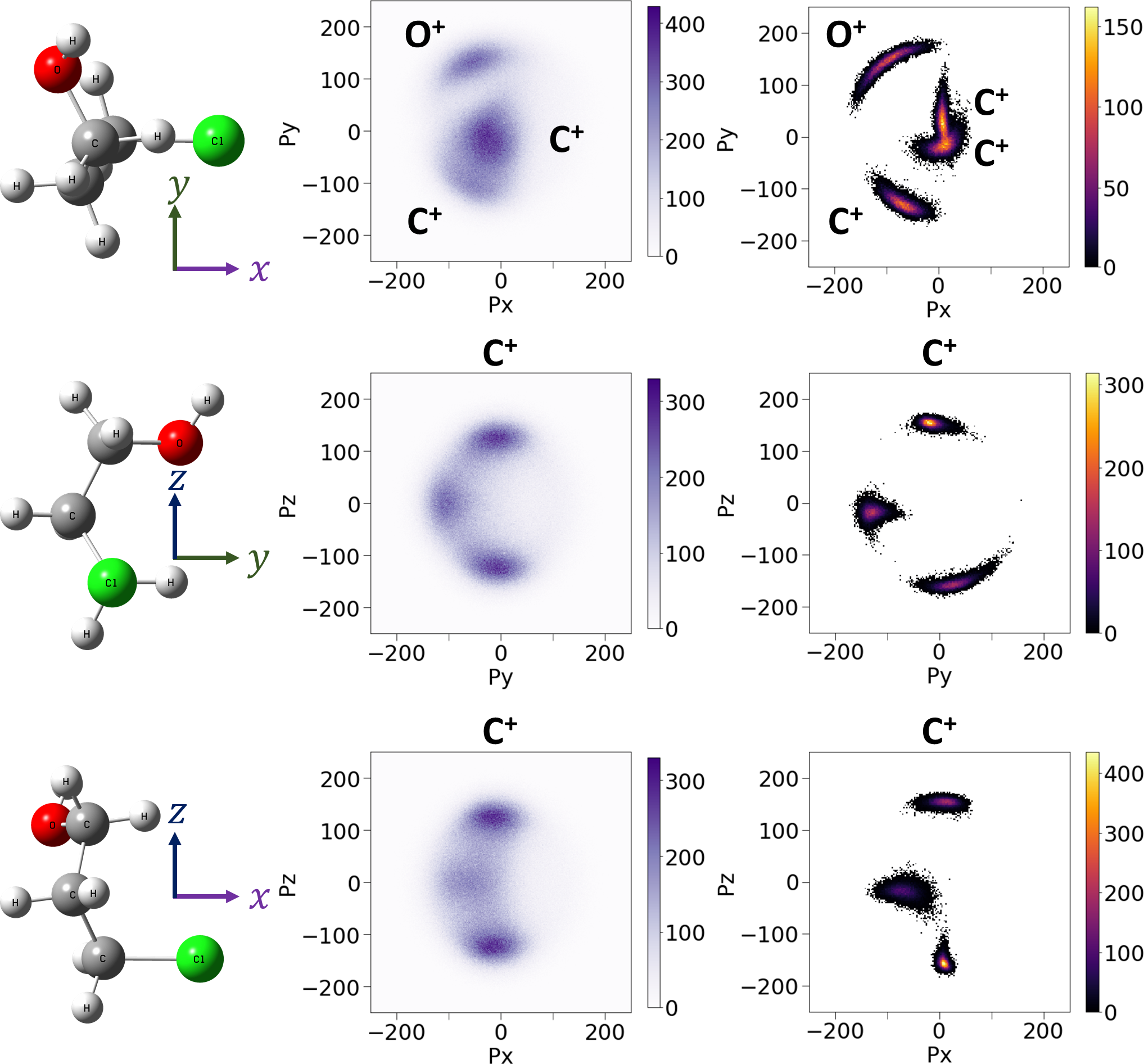}
  \caption{Left: molecular views. The ball-and-stick model of 3-chloro-1-propanol is rotated such that the C-Cl bond is along the $x$ axis, the C-O bond points in the positive $y$ direction, and the two carbons near Cl and O lie in the $xz$ plane. For this given rotation, all three projections are shown (top to bottom). Note that in the $zy$ projection, one carbon is hidden behind the chlorine. Center: projections of the 3D momenta in Fig.~\ref{fig:3d}(c) but not normalized (the momenta are in atomic unit). The momentum of $\mathrm{Cl^+}$ (not plotted) is along the $x$ axis. The $xy$-reference ion ($\mathrm{O^+}$) is plotted only in the top row (the $p_xp_y$ projection). Right: simulations of the projections in the middle panels using a classical Coulomb explosion model.}
  \label{fig:3chloro1propanol}
\end{figure}

To test the applicability of the CEI method to nonplanar molecules, Fig.~\ref{fig:3d} also shows the momentum images for 3-chloro-1-propanol, the open-chain molecule, in panel (b), and epichlorohydrin, the ring-chain molecule, in panel (c). Our experimental data show that CEI patterns of all three molecules exhibit distinguishable shapes with localized maxima. This opens up the possibility of distinguishing these structures if they are formed in a ring-opening reaction. This finding is particularly noteworthy considering that the open-chain and ring-chain structures are not as rigid as the structure of planar ring molecules. One might suspect that the momentum patterns would be much less structured due to possible motions and distortions during the explosion process \cite{Sanderson1999, Hishikawa1999, Hasegawa2001, Vager1993, Knoll2003, Lam2020, Comstock2003}, especially for molecules containing only single bonds, as investigated in this work. However, the comparison of the experimental data with the Coulomb explosion simulations in Fig.~\ref{fig:3chloro1propanol} shows that there is a very good correspondence between the fragment momentum correlations and the real-space molecular structure.

It is also worth noting that using a 3 kHz laser, these 3D structures can be obtained within one to a few hours. Figure~\ref{fig:3d} contains 6 hours of data on isoxazole and 3-chloro-1-propanol, and 1 hour of data on epichlorohydrin (see SM, Sec.~IV). This acquisition time can be significantly reduced since intense femtosecond lasers with much higher repetition rates (e.g, 100 kHz) are now commercially available. This makes it very promising and feasible to follow structural changes by tracking the motion of individual atoms \cite{Lam_coulomb_2023, Lam2023} of intermediate-size molecules of the order of ten atoms, making so-called molecular movies. Indeed, time-resolved CEI studies of the ultrafast UV-induced ring-opening dynamics of several molecules are currently ongoing in our lab \cite{Wang2023}.

A more detailed comparison between the experimental data, the real space geometry, and the simulation for a non-planar molecule, 3-chloro-1-propanol, is provided in Fig.~\ref{fig:3chloro1propanol}. Similar results for epichlorohydrin (the ring-chain molecule) can be found in the SM (Sec.~V). The experimental momentum images in Fig.~\ref{fig:3chloro1propanol} are obtained from the ($\mathrm{C^+}$, $\mathrm{C^+}$, $\mathrm{O^+}$, $\mathrm{Cl^+}$) 4-fold coincidence channel. The frame of reference is defined by the momentum vectors of $\mathrm{Cl^+}$ and $\mathrm{O^+}$ ($x$- and $xy$-reference ions, respectively); the $\mathrm{C^+}$ ions are plotted in this frame.

Figure 4 shows that even for nonplanar molecules with low symmetry, momentum distributions of the ions remain localized and separated, providing a clear individual correspondence between the real-space equilibrium molecular geometry and the momentum-space CEI patterns. Interestingly, the distinct curvatures exhibited by the three carbons in the $yz-$ and $xz-$planes in real space are also manifested in the CEI patterns in momentum space. Similar to the isoxazole case, simulations for 3-chloro-1-propanol capture the correlations between the fragment momenta qualitatively despite overestimating the momentum magnitudes. The experimental images exhibit diffuse stripes connecting the $\mathrm{C^+}$ momenta, which we attribute to vibrational excitation that happens during the strong-field ionization process (e.g., via Raman excitation). This can be reproduced in the simulation by increasing the temperature by a few hundred meV (SM, Sec. VI, Fig. 15). In all cases, the experimental patterns appear significantly broader than the simulations, indicating the potential influence of nuclear motions during the ionization and fragmentation process. The observed broadening may arise from sequential ionization involving multiple pathways converging to the same dissociation limit. Contribution from different final charge states can also broaden the distributions. These effects need further detailed investigations in future studies (e.g., investigating the effects of pulse duration on CEI, especially for hydrogen atoms \cite{Karimi2013, Tseng2018, Cheng2023}).

Open-chain and ring-chain structures have single bonds that are able to rotate, leading to the existence of multiple conformers with low energy barriers \cite{Fuller1982, Richardson1997, Badawi2008, Wang2000, Endo2020}. Given the strong laser fields used for CEI, it is remarkable how well a simple classical Coulomb explosion simulation model can capture the essential features of the experimental momentum images and point to the geometry of the neutral ground states (lowest energy conformers \cite{Fuller1982, Richardson1997, Badawi2008, Wang2000, Endo2020}). Our simulations show that other conformers, with similar energies but different geometries, give very distinct momentum patterns different from the experimental results (SM, Sec. VI, Fig. 14). This emphasizes the strength of CEI in distinguishing different conformer structures \cite{Pathak2020JPCL}. It also suggests that any structural changes occurring during an isomerization reaction will be reflected in time-resolved CEI patterns in momentum space.

In conclusion, we investigated the scalability of CEI using a kinematically incomplete coincidence analysis to image planar and nonplanar molecules with 8 or more atoms, in 3D momentum space, using intense tabletop lasers. Our result shows that momentum patterns of comparable clarity to recent results from XFELs \cite{Boll2022} can be obtained under these conditions. This opens the way for direct mapping of structural dynamics in all three dimensions. Since high-repetition-rate tabletop lasers are commercially available and can be easily implemented in a university-scale laboratory, laser-induced CEI offers a robust method for efficient structure identification and real-time imaging of photochemical reactions. While a simple classical Coulomb explosion model can qualitatively reproduce the experimental data, more advanced modeling and simulation \cite{Kwon1996, Ramadhan2017D, Sayler2018, Zhou2020, Richard2021, Minion2022} are highly desired to quantitatively retrieve the molecular structures in real space (i.e., bond lengths and bond angles) from the momentum space measurements.

~
\begin{acknowledgments}
We acknowledge Shashank Pathak and S.~Javad Robatjazi for commissioning the double-sided velocity map imaging apparatus. We are grateful to the technical staff of the J.R.~Macdonald Laboratory for their tireless support and for the complete remodel of our laser lab. We thank the Computational Physics Key Laboratory, Department of Physics, Ho Chi Minh City University of Education, Ho Chi Minh City, Vietnam for providing certain resources used in the simulation. This work, the procurement of the laser system, and the operation of the J.R.~Macdonald Laboratory are supported by the Chemical Sciences, Geosciences, and Biosciences Division, Office of Basic Energy Sciences, Office of Science, U.S. Department of Energy, Grant no.~DE-FG02-86ER13491.  S. B. is supported by the U.S. Department of Energy Established Program to Stimulate Competitive Research (DOE-EPSCoR) Grant no. DE-SC0020276. A. S. V. is supported by the National Science Foundation Grant no. PHYS-1753324.
\end{acknowledgments}



\bibliography{CEI_single_pulse_2022}

\typeout{get arXiv to do 4 passes: Label(s) may have changed. Rerun}

\end{document}


\preprint{}

\title{Supplementary Material for\\
Differentiating Three-Dimensional Molecular Structures using\\
Laser-induced Coulomb Explosion Imaging}

\author{Huynh Van Sa Lam}
\email{\textcolor{black}{huynhlam@phys.ksu.edu}}
\affiliation{James R. Macdonald Laboratory, Kansas State University, Manhattan, KS 66506, USA}
\author{Anbu Selvam Venkatachalam}
\affiliation{James R. Macdonald Laboratory, Kansas State University, Manhattan, KS 66506, USA}
\author{Surjendu Bhattacharyya}
\affiliation{James R. Macdonald Laboratory, Kansas State University, Manhattan, KS 66506, USA}
\author{Keyu Chen}
\affiliation{James R. Macdonald Laboratory, Kansas State University, Manhattan, KS 66506, USA}
\author{Kurtis Borne}
\affiliation{James R. Macdonald Laboratory, Kansas State University, Manhattan, KS 66506, USA}
\author{Enliang Wang}
\affiliation{James R. Macdonald Laboratory, Kansas State University, Manhattan, KS 66506, USA}
\author{Rebecca Boll}
\affiliation{European XFEL, 22869 Schenefeld, Germany}
\author{Till Jahnke}
\affiliation{European XFEL, 22869 Schenefeld, Germany}
\author{Vinod Kumarappan}
\affiliation{James R. Macdonald Laboratory, Kansas State University, Manhattan, KS 66506, USA}
\author{Artem Rudenko}
\affiliation{James R. Macdonald Laboratory, Kansas State University, Manhattan, KS 66506, USA}
\author{Daniel Rolles}
\email{\textcolor{black}{rolles@phys.ksu.edu}}
\affiliation{James R. Macdonald Laboratory, Kansas State University, Manhattan, KS 66506, USA}


\date{\today}

\maketitle

\tableofcontents

\clearpage

\section{\label{sec:method} Experimental setup}

\begin{figure}[h]
\includegraphics[width=0.6\columnwidth]{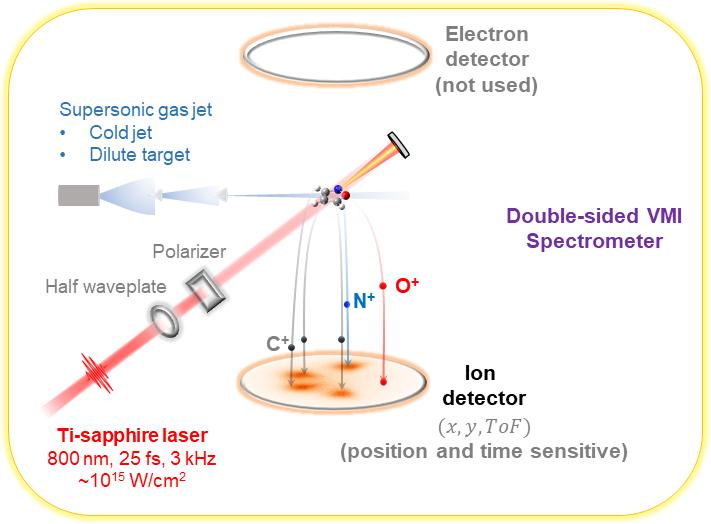}
  \caption{Schematic of the experimental setup.}
  \label{fig:expt_setup}
\end{figure}

Figure~\ref{fig:expt_setup} shows a schematic of the experimental setup consisting of a double-sided coincidence velocity-map imaging (VMI) spectrometer, a doubly skimmed supersonic molecular beam, and a 3-kHz femtosecond near-infrared laser (Coherent Legend Elite DUO) with a central wavelength of 810 nm. The laser polarization is horizontal (linear and parallel to the detector surface). The laser pulse energy was controlled by using a combination of an achromatic zero-order half-wave plate and a thin-film polarizer. The pulse duration was measured using a frequency-resolved optical gating (FROG) \cite{Trebino1997} before the beam entered the vacuum chamber. In order to account for dispersion in the vacuum chamber entrance window (1 mm CaF$_2$), an identical window was added before the FROG setup. The compressor grating was optimized for the shortest pulse ($\approx$~25~fs). The pulses were focused by an in-vacuum 75-mm focal length back-reflecting concave mirror into the cold, dilute molecular samples produced by supersonic expansion of the respective sample molecule and helium carrier gas at a total pressure of $\approx$~20 psi through a 30-$\mathrm{\mu}$m nozzle. A 500-$\mathrm{\mu}$m skimmer is placed within a few millimeters after the nozzle to skim the center of the expanding gas beam. In this experiment, only the ion side of the VMI spectrometer was used.

\begin{figure}[h]
\includegraphics[width=0.7\columnwidth]{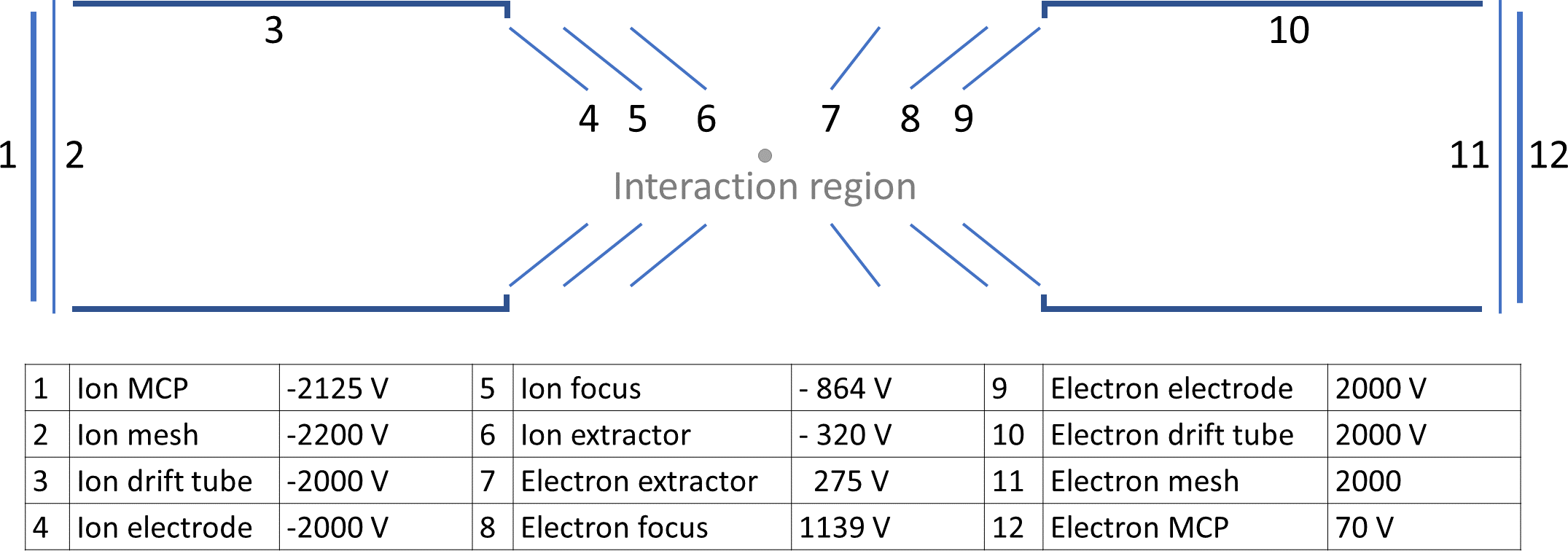}
  \caption{A sketch of the double-sided VMI spectrometer and a table of voltages used in the experiment.}
  \label{fig:spectrometer}
\end{figure}

Figure~\ref{fig:spectrometer} shows a sketch of the spectrometer and its operating voltages, while more detailed descriptions can be found in Refs.~\cite{Robatjazi2021D, Pathak2021D}. The detector on the ion side consists of two microchannel plates (MCP) with 80-mm diameter (one funnel MCP on the entrance side and one typical MCP with constant pore size at the back) followed by a Roentdek DLD-80 delay-line anode. The fragment ions were detected in coincidence, and their momenta were determined from the measured time-of-flight (ToF) and impact position of each ion. The total ion count rate was about 1.9 kHz for 3-chloro-1-propanol, 5 kHz for isoxazole and 6.5 kHz for epichlorohydrin.

\section{\label{sec:CCES} Classical Coulomb explosion simulation}

\subsection{Description of the model}

In the classical Coulomb explosion simulations for the equilibrium geometry, we first use the Gaussian software package \cite{g09} to optimize the geometry of the neutral molecule in its electronic ground state at the B3LYP/aug-cc-pvdz level and calculate vibrational modes and frequencies. After that, we use the NewtonX software \cite{Barbatti2014, NTX} to sample 20,000 geometries of the initial distribution in the ground state at 300 K.

The INITCOND function from NewtonX generates initial conditions based on normal modes by sampling the amplitude and momentum of quantum mechanical harmonic oscillator distributions. We use NACT = 2 where the distribution matches the Wigner distribution for the quantum harmonic oscillator if the vibrational numbers are zero (ground vibrational state). For higher vibrational quantum numbers, the distribution for each normal mode is a product of the harmonic oscillator wavefunctions in the coordinate and momentum spaces. The sampling of the coordinates and momenta is uncorrelated. At a temperature T (K), the Wigner sampling is broadened by a factor of $\mathrm{\tanh \left( h\omega/2k_BT \right)}$.

Finally, we perform classical Coulomb explosion simulations (CES) for the sampled geometries by numerically solving Newton's equations of motion for eight point charges initiated at the position of each atom in the molecule. This simple model assumes that all 4-fold coincident events stem from an 8-fold ionization of the molecule, and that the repulsive potential of the highly-charged cations leading to multibody fragmentations is purely Coulombic. It also assumes that the charge distribution of the cation can be described as an array of instantaneously-created point charges, and that the molecule fragments completely into eight singly charged atomic ions without any internal energy.

\subsection{Comparision with experiment and limitation of the model}

\begin{figure}[h]
    \includegraphics[width=\columnwidth]{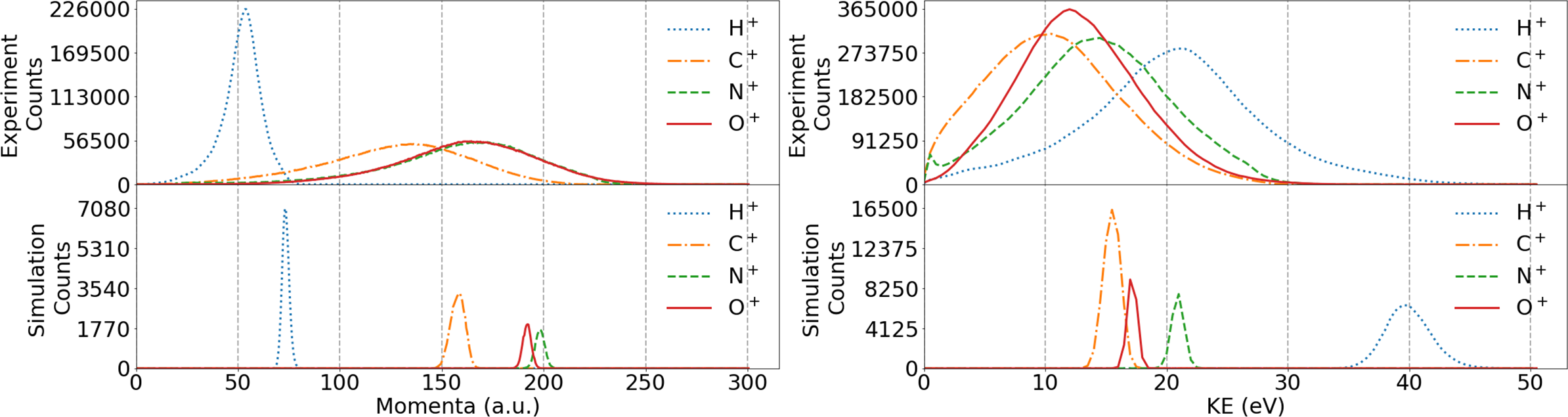}
    \caption{Comparison between the measured (top) and the simulated (bottom) momentum magnitude (left) and kinetic energy (right) of different fragments in Coulomb explosion imaging of isoxazole. The measured data is obtained from the 4-fold ($\mathrm{H^+,C^+,N^+,O^+}$) coincidence channel. The simulated momentum distributions for carbon ions and protons include all three carbon ions and three protons.}
    \label{fig:isoxazole_momenta}
\end{figure}

\begin{figure}
    \includegraphics[width=0.8\columnwidth]{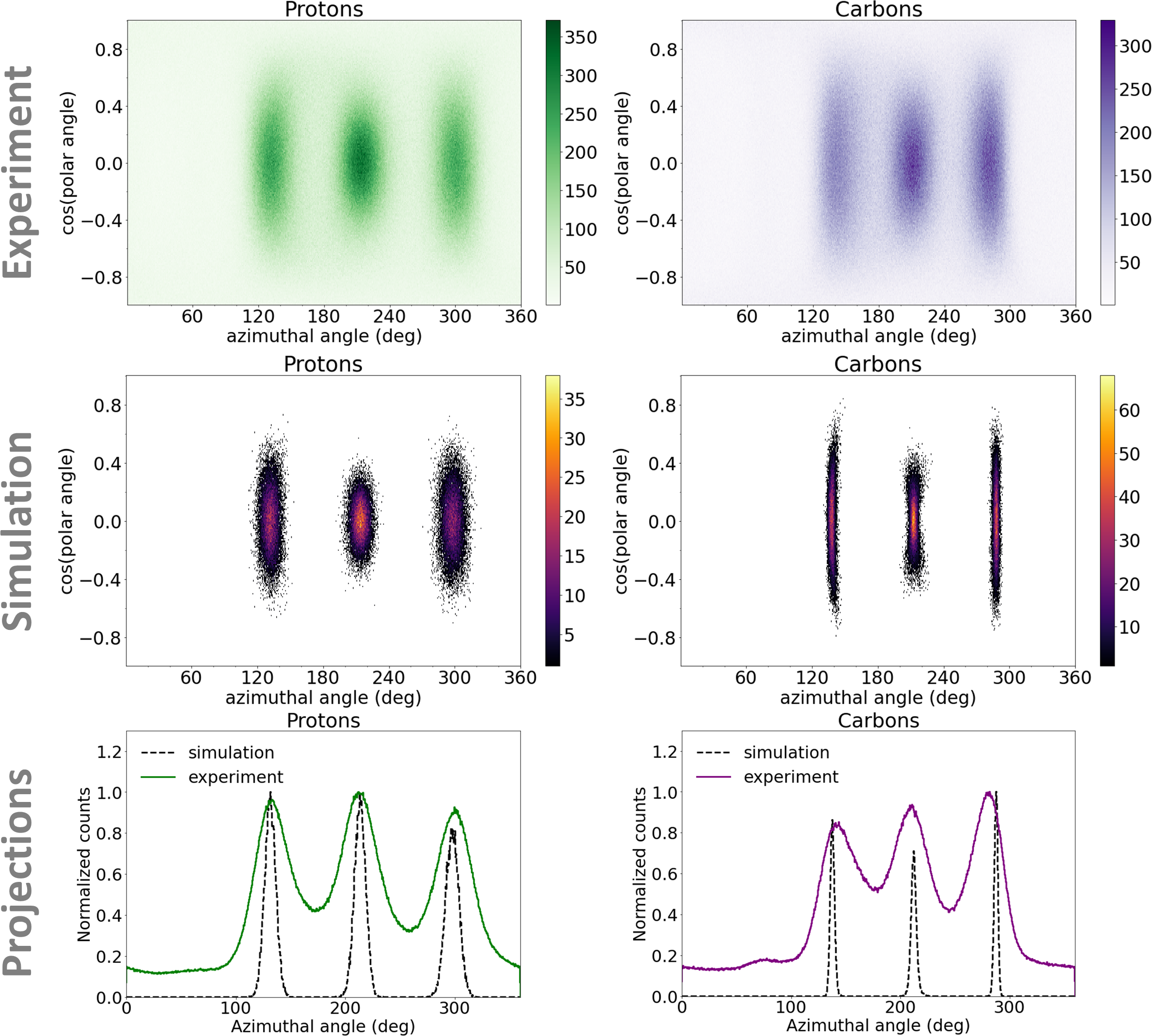}
    \caption{Angle correlations between momentum vectors of different fragments of the 4-fold ($\mathrm{H^+,C^+,N^+,O^+}$) coincidence channel. The left panel shows the results for protons, and the right panel shows the results for carbon ions. The top two rows show 2D histograms where the horizontal axis is the azimuthal angle in the $p_xp_y$-projection and the vertical axis is the cosine of the polar angle obtained from the experimental data (top) and simulation (middle). The bottom row shows the projections of both experiment and simulation data onto the $x$-axis.}
    \label{fig:isoxazole_angles}
\end{figure}

Although this model overestimates the momentum magnitudes of the fragments, it captures the correlations between the fragment momenta rather accurately, as shown in Figs.~\ref{fig:isoxazole_momenta} and \ref{fig:isoxazole_angles}.
Figure~\ref{fig:isoxazole_momenta} shows a comparison between the magnitude of the simulated and the experimental momenta and kinetic energies (KEs), while Fig.~\ref{fig:isoxazole_angles} compares the angular correlations between momentum vectors of different fragments of the 4-fold ($\mathrm{H^+,C^+,N^+,O^+}$) coincidence channel. These results demonstrate that although our classical Coulomb explosion model tends to overestimate the momentum magnitudes (or KEs) of the fragments, it captures the correlations between the fragment momenta (i.e., the angles between momentum vectors) rather accurately.

In general, the assumption of instantaneous ionization is most problematic for light atoms (such as, in particular, hydrogens), which move faster than the heavier atoms and, therefore, have time to move the most during the ionization and dissociation process. Furthermore, the larger the molecules, the higher charge states are required to completely explode the molecule into its atomic fragments, which typically means that the ionization process is extended over a longer time. We therefore suspect that the complications for imaging hydrogens might increase with the size of the molecule, although more systematic studies are required to verify this expectation.

More elaborate Coulomb explosion models include a “charge build-up model” that takes into account the time it takes for the charges to build up in the molecule rather than instantaneously placing a positive charge on all atoms. For the case of x-ray-induced ionization, this can be done on the basis of \textit{ab initio} modeling of the ionization process, as done in Boll \textit{et al.} (2022) \cite{Boll2022}, or via some empirical model that is fitted to the data [see, for instance, Motomura \textit{et al.} (2015) \cite{Motomura2015} and Li \textit{et al.} (2022) \cite{Li2022}. The latter could also be extended to the case of strong-field ionization. For the present experiment, our instantaneous model is sufficient to reproduce the key features observed in the experimentally determined momentum correlations, but a charge-buildup model might be able to improve the quantitative agreement between experimental and simulated kinetic energies, for example.\\

\vspace{30pt}

\clearpage
\section{\label{sec:coinc_channels} Additional results on isoxazole}

\subsection{Comparison between different coincidence channels}

\begin{figure}[h]
    \includegraphics[width=\columnwidth]{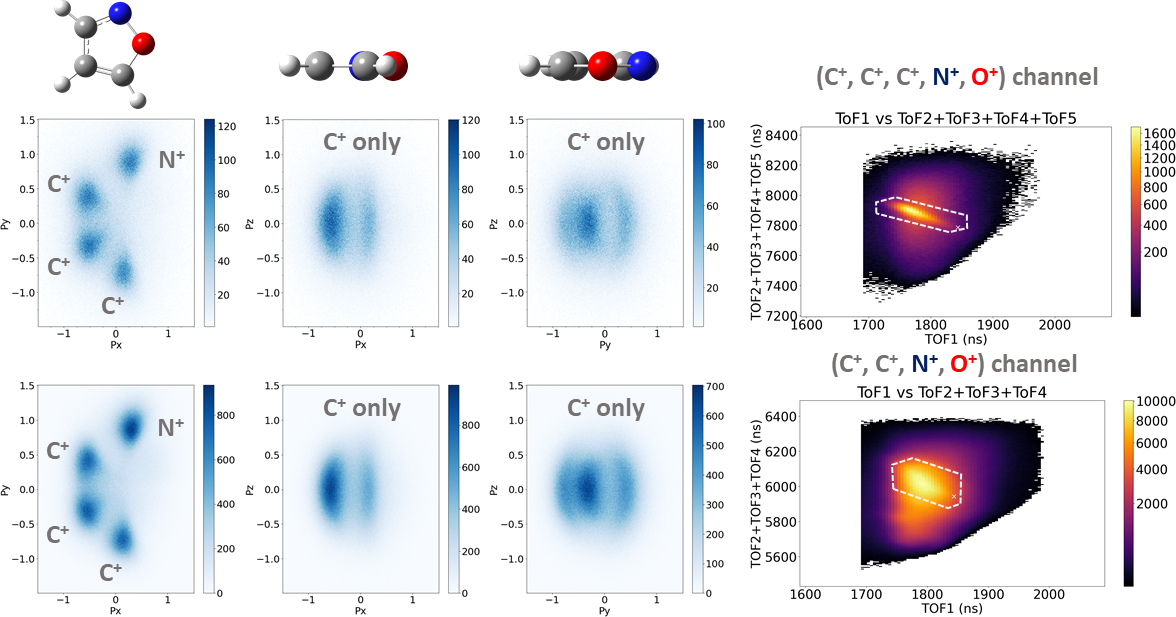}
    \caption{Comparison between the CEI patterns obtained from the 5-fold ($\mathrm{C^+,C^+,C^+,N^+,O^+}$; top row) and the 4-fold ($\mathrm{C^+,C^+,N^+,O^+}$; bottom row) coincidence channels.}
    \label{fig:isoxazole_4_vs_5_body}
\end{figure}

\begin{figure}[h]
    \includegraphics[width=0.8\columnwidth]{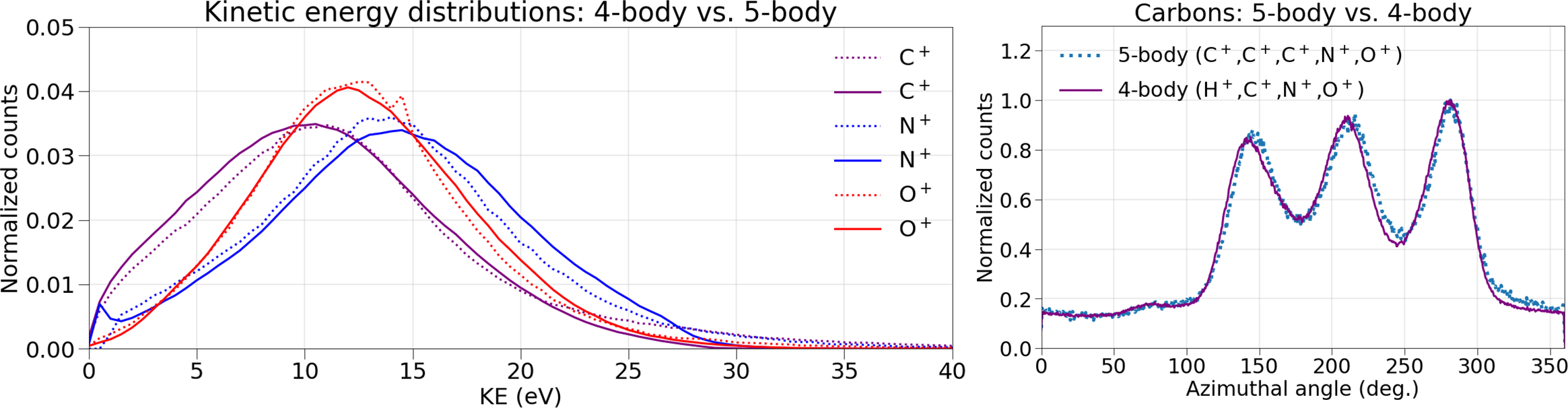}
    \caption{Comparison between the kinetic energy (left) and azimuthal angle (right) distributions obtained from the 4-fold ($\mathrm{H^+,C^+,N^+,O^+}$) (solid) and the 5-fold ($\mathrm{C^+,C^+,C^+,N^+,O^+}$) (dotted) coincidence channels. The kinetic energy distribution of each fragment is normalized by its area under the curve. The distribution of azimuthal angle is normalized to its maximum value. For the 5-fold ($\mathrm{C^+,C^+,C^+,N^+,O^+}$) coincidence channel, the distributions of ($\mathrm{C^+}$) include all the detected carbon ions.}
    \label{fig:isoxazole_4_vs_5_quan}
\end{figure}

Figure~\ref{fig:isoxazole_4_vs_5_body} shows a comparison between the CEI patterns obtained from 5-fold ($\mathrm{C^+,C^+,C^+,N^+,O^+}$) (top) and 4-fold ($\mathrm{C^+,C^+,N^+,O^+}$) (bottom) coincidence channels of isoxazole from the same data set presented in the main text. The statistics of the former is about a factor of 10 lower than that of the 4-fold coincidence channels, but the shape and clarity of the momentum images are similar to the 4-fold ($\mathrm{H^+,C^+,N^+,O^+}$) coincidence channel presented in the main text (see also a more quantitative comparison in Fig.~\ref{fig:isoxazole_4_vs_5_quan}).

In Fig.~\ref{fig:isoxazole_4_vs_5_quan}, we compare the 4-fold ($\mathrm{H^+,C^+,N^+,O^+}$) (solid) and the 5-fold ($\mathrm{C^+,C^+,C^+,N^+,O^+}$) (dotted) coincidence channels in terms of the kinetic energy of $\mathrm{C^+}$, $\mathrm{N^+}$, and $\mathrm{O^+}$ fragments (left) and the distribution of the azimuthal angle (right). 
The very close agreement suggests that the ions in the 4-fold and the 5-fold coincidence channels most likely stem from the same final charge state of the molecule.

\subsection{Discussion on momentum gating}

The right panel of Fig.~\ref{fig:isoxazole_4_vs_5_body} shows the ToF-coincidence maps, where the $x$-axis is the ToF of the first ion (first $\mathrm{C^+}$) and the $y$-axis is the total ToFs of all the other fragments. Note that the TOF-coincidence map of the 5-fold ($\mathrm{C^+,C^+,C^+,N^+,O^+}$) channel shows a reasonably sharp line, indicating that the momentum sum of this channel is close to a constant value and that the contribution of the hydrogen fragments to the total momentum is small. In this case, we can set a momentum gate (as indicated by the white contour) on these coincidence events and thus strongly suppress the contribution of false coincidences. Such a momentum gate was applied in the 5-fold coincidence data presented in  Fig.~\ref{fig:isoxazole_4_vs_5_body}.

\begin{figure}[h]
    \includegraphics[width=0.8\columnwidth]{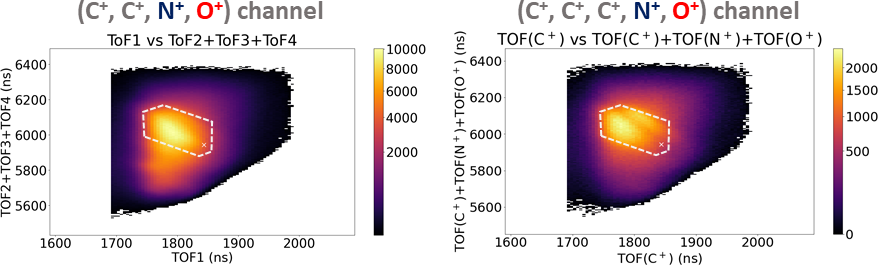}
    \caption{Comparison between the coincidence maps obtained from the the 4-fold ($\mathrm{C^+,C^+,N^+,O^+}$) (left) and the 5-fold ($\mathrm{C^+,C^+,C^+,N^+,O^+}$) (right) coincidence channels where only two out of three $\mathrm{C^+}$ ions are used. One carbon is used to define TOF($\mathrm{C^+}$) on the $x$-axis while the other carbon is used as TOF($\mathrm{C^+)}$ in the sum TOF($\mathrm{C^+}$)+TOF($\mathrm{N^+}$)+TOF($\mathrm{O^+}$) on the $y$-axis. The combinations are C(1)C(2), C(1)C(3) and C(2)C(3) where the numbers indicate the order the carbon was detected. The coincidence map on the right looks like it contains two separate islands. The top-right feature comes from the C(2)C(3) combination since both carbon ions have longer ToFs than C(1). The white contour is the gate used for selecting events of the 4-fold ($\mathrm{C^+,C^+,N^+,O^+}$) coincidence channel. It is overlaid on the right coincidence map made from the 5-fold channel. This gives us an idea of how to select events from the 4-fold channel.}
    \label{fig:isoxazole_4_vs_5_coinc}
\end{figure}

In the 4-fold coincidence channels, there is no sharp line since the missing heavy fragments carry non-negligible momentum, such that a strict momentum gating is not possible. In this case, we gated on the most intense spot of the coincidence map. To ascertain this selection, we use two out of three carbons from the 5-fold coincidence channel to make a coincidence map containing four ions, as shown in Fig.~\ref{fig:isoxazole_4_vs_5_coinc}(b). This map gives us an idea of where the 4-fold ($\mathrm{C^+,C^+,N^+,O^+}$) channel will appear in the coincidence map. The area suggested by this method largely coincides with the most intense spot identified in the coincidence map in Fig.~\ref{fig:isoxazole_4_vs_5_body}.

The importance of the momentum gating varies for each molecule and coincidence channel. Isoxazole is a difficult case since all the fragment ions (i.e., C$^+$, N$^+$, O$^+$) are close in time of flight, which causes a substantial overlap between heavier fragments emitted towards the detector and lighter fragments emitted away from the detector. They are also easily contaminated by ions from the background, such as air and water. This can lead to a mis-assignment of the ion species and some artifacts in determining the fragment momenta. To a large extent, this can be avoided by carefully selecting appropriately narrow momentum gates as described above. To illustrate this point, we produce the Newton plots for isoxazole with and without the momentum gates from the 4-fold  ($\mathrm{C^+,C^+,N^+,O^+}$) coincidence channel in Fig.~\ref{fig:isoxazole_mom_gate}.

\begin{figure}[h]
    \includegraphics[width=0.5\columnwidth]{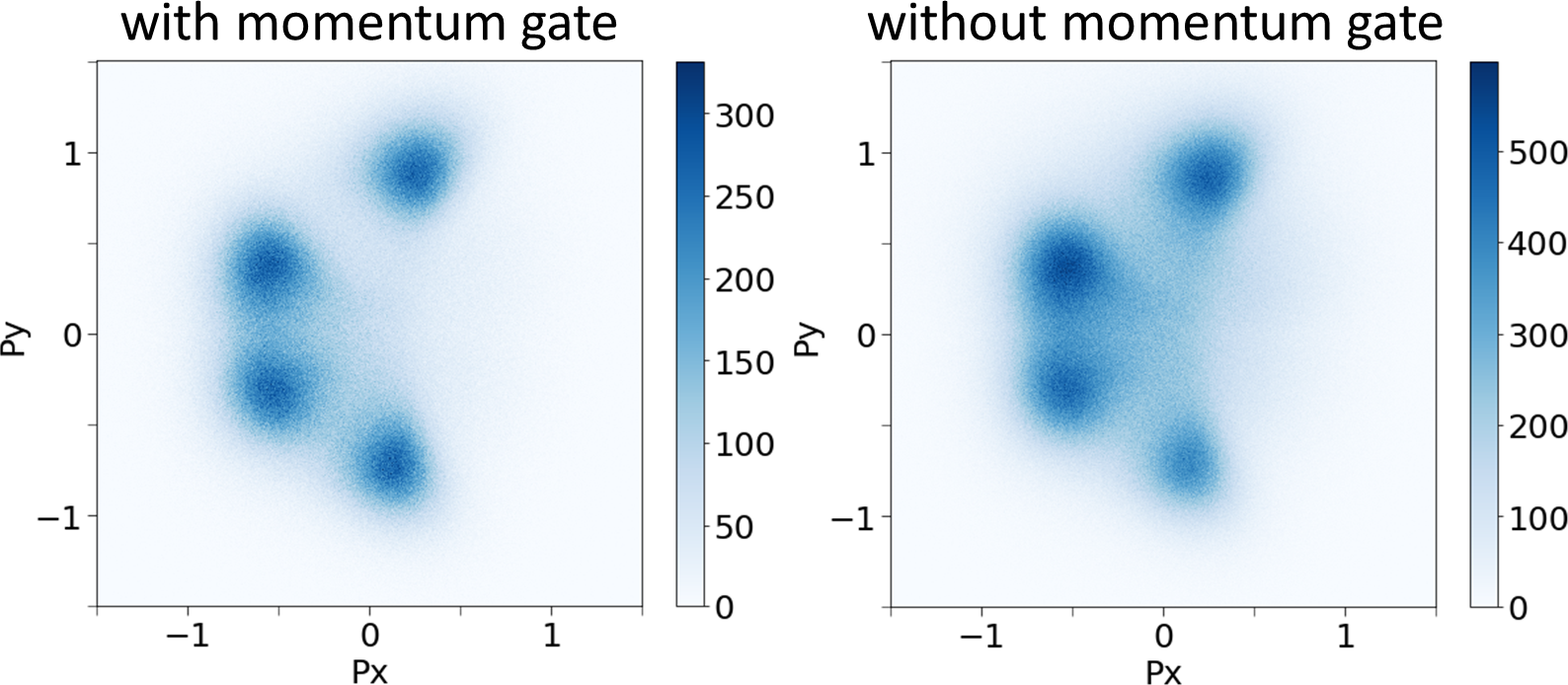}
    \caption{Comparison between the momentum images obtained from the 4-fold ($\mathrm{C^+,C^+,N^+,O^+}$) coincidence channel with and without the momentum gate indicated by the white contour in Fig.~\ref{fig:isoxazole_4_vs_5_body}.}
    \label{fig:isoxazole_mom_gate}
\end{figure}

The pattern in the plot without the momentum gate is more smeared out, but all the main structures can still be identified well. Therefore, this gating is more important in pump-probe experiments where the pump-induced signal is typically only a fraction of the probe-only signal and can be easily obscured by smeared-out structures resulting from the lack of proper gating. However, it is worth noting that, in many CEI analyses employing kinematically incomplete channels, such momentum gating is not possible due to the difficulty of detecting all the C$^+$ and H$^+$ ions to narrow down the momentum gate.

For example, as shown in the main text, the 4-fold ($\mathrm{H^+,C^+,N^+,O^+}$) coincidence channel allows us to construct momentum images corresponding to all eight atoms in the isoxazole molecule. However, false events will dominate the signal since only one H$^+$ and one C$^+$ were detected, and no strict momentum gating is available. In this case, we use an alternative approach that is less reliant on the choice of appropriate momentum gates. We discard all events in the ambiguous time-of-flight regions by only using the right half of the time-of-flight peaks of N$^+$ and O$^+$ to avoid the overlap between C$^+$, N$^+$, and O$^+$ ions. We also discard events in the regions where contamination from background residual gas molecules (mostly water and air) arises (see Fig.~\ref{fig:tof_isoxazole} in Sec.~\ref{TOF} for ToF and position-ToF spectra).

In Fig.~\ref{fig:isoxazole_mom_gate} below, we produce the coincidence maps and the Newton plots for isoxazole from the 4-fold ($\mathrm{H^+,C^+,N^+,O^+}$) coincidence channel with and without the gates described above. The coincidence maps show that only certain events above the black cross were selected, while most events in other regions (including the most intense spot) were discarded. This is because we keep N$^+$ and O$^+$ ions with larger time-of-flights, leading to a higher sum on the vertical axis. In the lower panel, the momentum image with gating is much cleaner, showing clear structures, while very limited structural information can be observed from the momentum image without gating. As elucidated in Sec.~\ref{sec:coinc_channels} A above, our results demonstrate that both approaches yield clean CEI patterns of comparable quality, irrespective of whether one, two, or all three C$^+$ ions were detected using different coincidence channels.

\begin{figure*}[h]
    \includegraphics[width=\columnwidth]{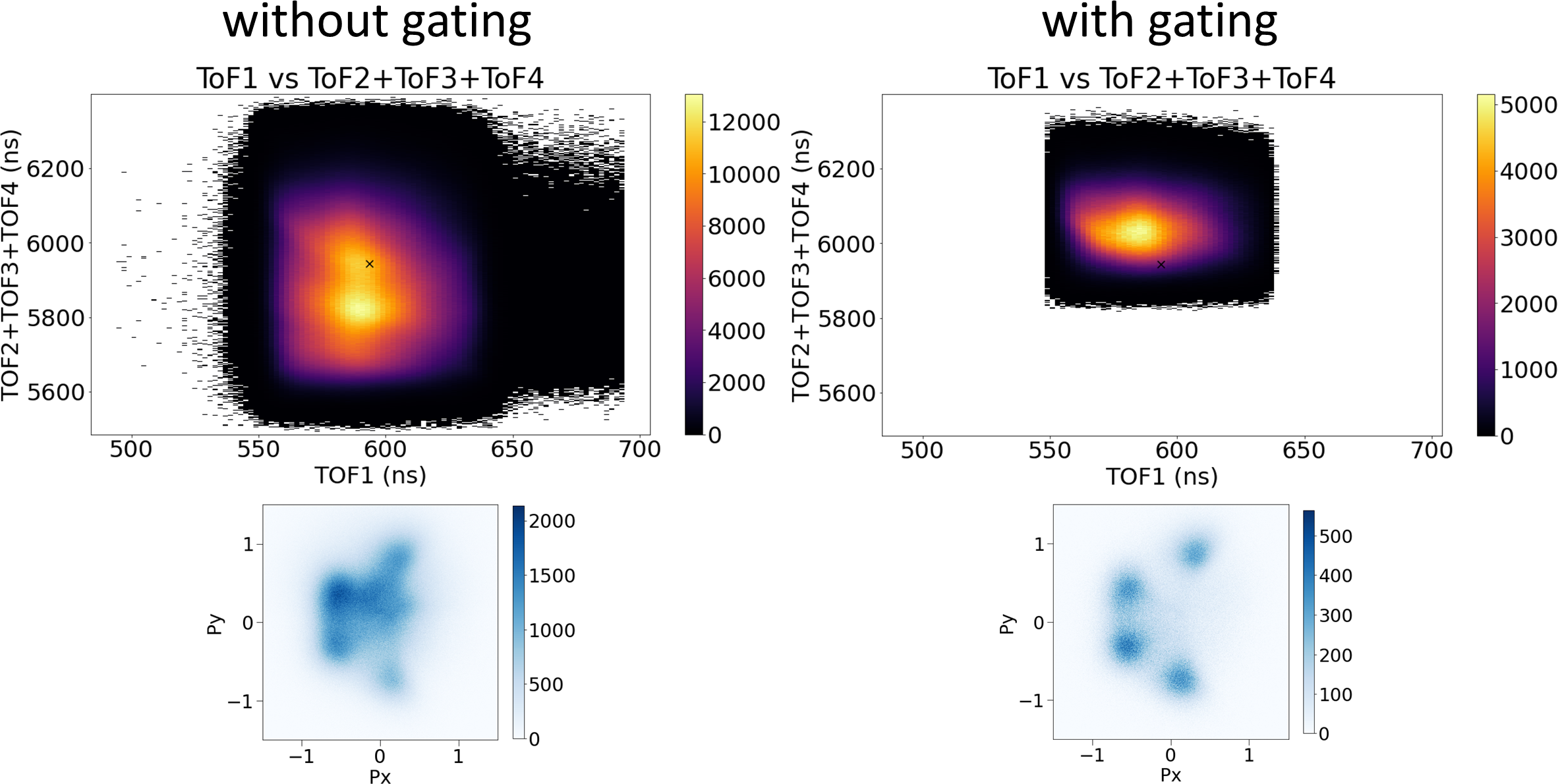}
    \caption{Coincidence maps and the Newton plots for isoxazole from the 4-fold ($\mathrm{H^+,C^+,N^+,O^+}$) coincidence channel with and without gating (see text for details).}
    \label{fig:isoxazole_mom_gate}
\end{figure*}

\subsection{Discussion on the out-of-plane momentum distributions of the carbon ions}

Our data and simulations also capture the appearance of the middle carbon in the $yz$-plane (as compared to the $xz$-plane) with a smaller out-of-plane momentum spread than the other two carbons (as seen in Fig.~\ref{fig:isoxazole_4_vs_5_body} here and also in the main text). A more detailed analysis below concludes that this different visual appearance of the different atoms reflects different degrees of correlated motion with respect to the reference atoms(s). To discuss this for a specific example, we consider why the “middle carbon”, labeled C(2) in the figure below, appears to have a smaller out-of-plane momentum spread than the other two carbons, labeled C(1) and C(3). It turns out that this observation strongly depends on the reference frame that is chosen for the plot, which can be easily varied in our Coulomb explosion simulations.

In panel (a) of Fig.~\ref{fig:corr_motions} below, we choose the molecular plane based on the planar equilibrium geometry of the molecule and allow each atom to freely move “in-plane” and “out-of-plane” according to the normal modes of the molecule. For each geometry in this ensemble of “vibrating” molecules, we perform a Coulomb explosion simulation and then plot the p$_z$ component of the momentum vectors of the carbon ions with respect to the original molecular plane. In this representation, all three carbon ions have the same out-of-plane momentum spread.

In panel (b), we produce a similar plot, but the molecular plane is now defined by the momentum vectors of O$^+$ and N$^+$ on an event-by-event basis, as we do in the experiment. In this case, the out-of-plane momentum distribution of the middle carbon C(2) is comparable to the original distribution, while the distributions of the other two carbons [C(1) and C(3)] become broader compared to the left panel. This reflects the fact that, as a result of the normal modes, the motion of the middle carbon is correlated with the motion of the nitrogen and oxygen atoms, and thus, its momentum vector (p$_z$) is more likely to lie closer to the plane defined by p(N$^+$) and p(O$^+$).

\begin{figure}[h]
    \includegraphics[width=\columnwidth]{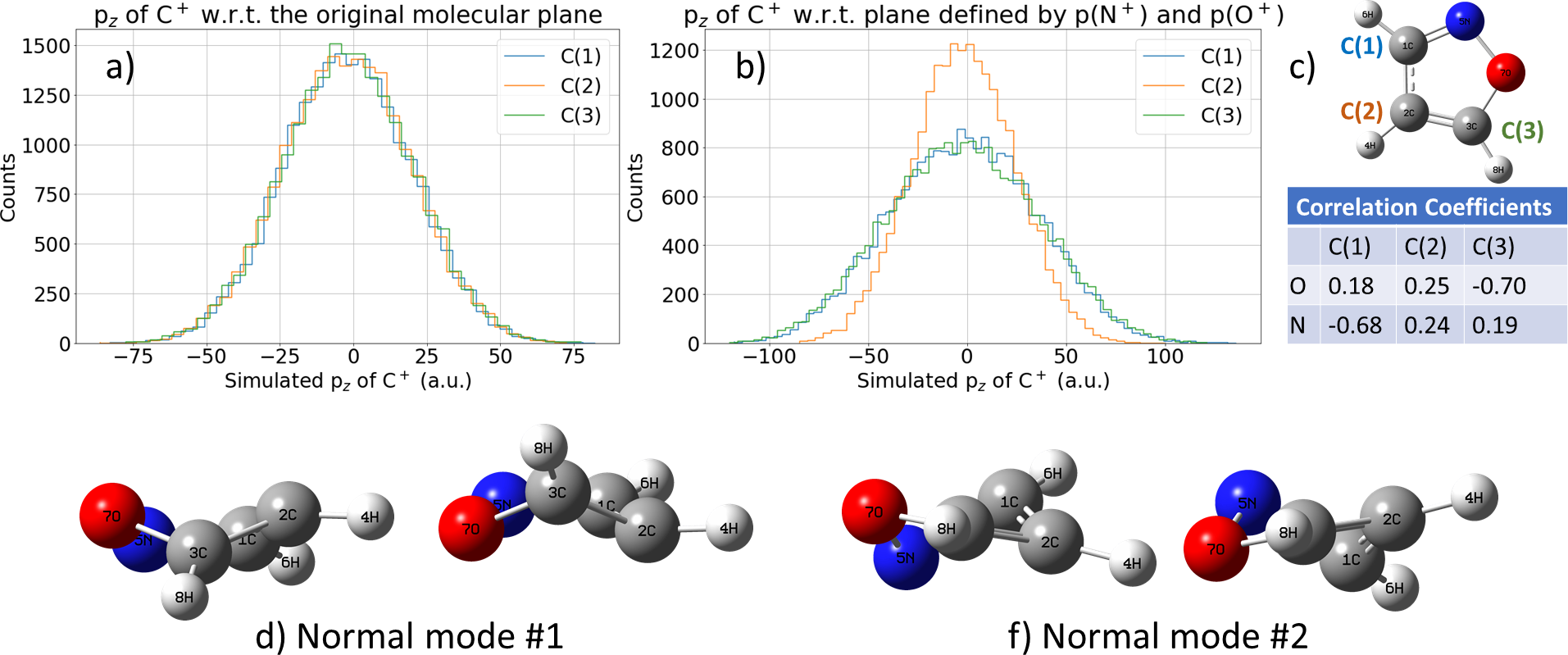}
    \caption{Correlations between different carbon ions with N$^+$ and O$^+$, see text for more details.}
    \label{fig:corr_motions}
\end{figure}

To quantify this correlated motion, we calculated the correlation coefficient between p$_z$(C$^+$) and p$_z$(N$^+$) or p$_z$(O$^+$), shown in panel (c) of Fig.~\ref{fig:corr_motions} (p$_z$ is with respect to the planar equilibrium geometry). C(2) has small positive correlations, meaning its p$_z$ is more likely to vary in the same direction as p$_z$(N$^+$) or p$_z$(O$^+$). On the other hand, C(1) and C(3) show strong negative correlations with either p$_z$(N$^+$) or p$_z$(O$^+$), meaning their p$_z$ is more likely to vary in the opposite direction as p$_z$(N$^+$) or p$_z$(O$^+$. This simple analysis suggests that when p(N+) and p(O+) are chosen to define the “molecular” plane, the out-of-plane momentum for C(1) and C(3) could thus be broadened to a larger magnitude as compared to C(2).

An investigation of normal modes shows two normal modes that exhibit strong out-of-plane motion of N and O, as shown in Fig.~\ref{fig:corr_motions}(c,d). In the first normal mode, the middle carbon C(2) moves in the same direction as N and O, which leads to a similar variation of their p$_z$ components. In the second normal mode, C(2) moves in the same direction as N, but in the opposite direction with O, so we expect the effect to cancel out to some extent.

This analysis suggests that even data taken with a \textit{single} IR pulse, i.e., not in a pump-probe scheme, contains some information about the correlated motions between different atoms in the molecule. This is an interesting avenue; however, further investigations, which are beyond the scope of the current manuscript, are required to explore these correlations in more detail.

\clearpage
\section{\label{sec:acq_time} Buildup of 3D structures as a function of data acquisition time}

\begin{figure}[h]
    \includegraphics[width=0.8\columnwidth]{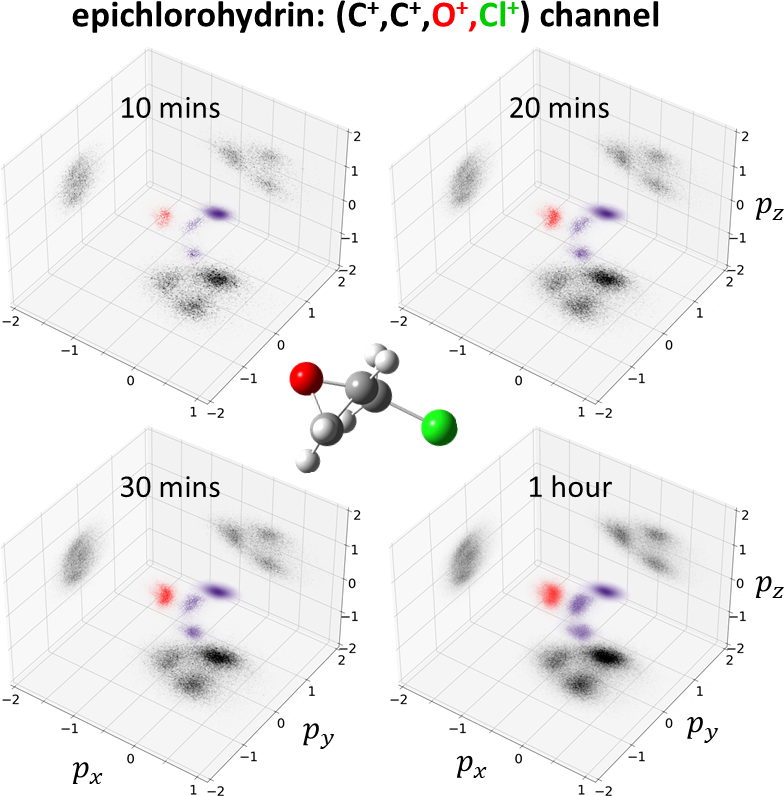}
    \caption{3D momentum images of epichlorohydrin as a function of data acquisition time. The sample has a vapor pressure of approximately 17 mmHg at 25$^\circ$C, and the ion count rate was approximately 6.5 kHz (at a laser repetition rate of 3 kHz). On average, we obtained $\approx$~16900 events of the 4-fold ($\mathrm{C^+,C^+,N^+,O^+}$) coincidence every 10 minutes.}
    \label{fig:epichlorohydrin_3D_time}
\end{figure}

\begin{figure}[h]
    \includegraphics[width=0.8\columnwidth]{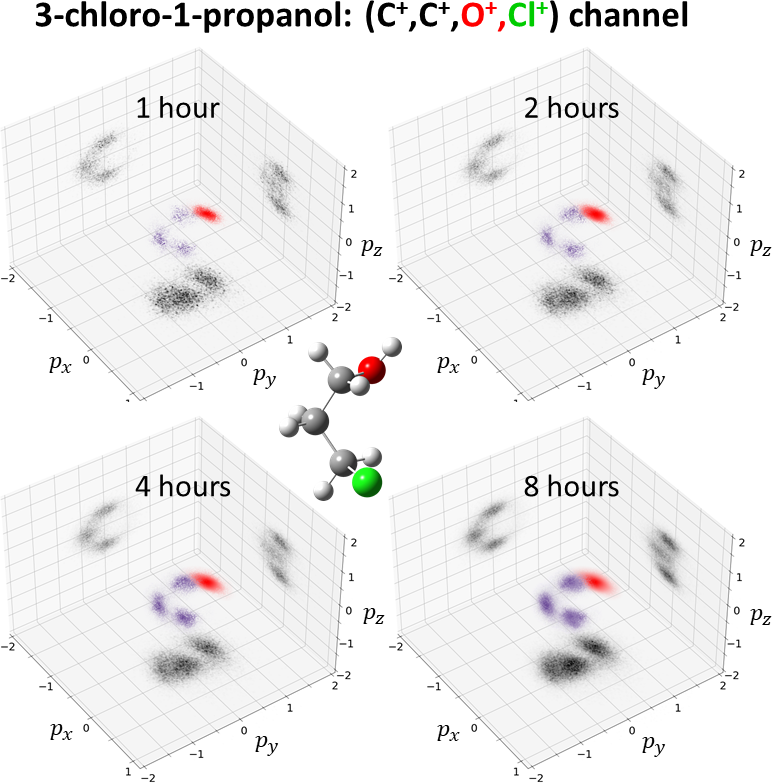}
    \caption{3D momentum images of 3-chloro-1-propanol as a function of data acquisition time. The sample has a relatively low vapor pressure of approximately 1 mmHg at 25$^\circ$C, and the ion count rate was approximately 1.9 kHz (at a laser repetition rate of 3 kHz). On average, we obtained $\approx$~8300 events of the 4-fold ($\mathrm{C^+,C^+,N^+,O^+}$) coincidence every hour.}
    \label{fig:3_chloro_1_propanol_3D_time}
\end{figure}

Figure~\ref{fig:epichlorohydrin_3D_time} and Fig.~\ref{fig:3_chloro_1_propanol_3D_time} show the buildup of the measured 3D structures in momentum space as a function of data acquisition time. Using a 3 kHz laser, clear 3D structures can be obtained within a few tens of minutes (for the epichlorohydrin data collected at an ion count rate of 6.5 kHz) or a few hours (for 3-chloro-1-propanol data collected at a lower ion count rate of 1.9 kHz due to the lower vapor pressure of the sample).  For the 4-fold ($\mathrm{C^+,C^+,N^+,O^+}$) coincidence channel, on average, we obtained $\approx$~8300 events every hour for 3-chloro-1-propanol and $\approx$~101500 events every hour for epichlorohydrin. Commercially-available intense femtosecond lasers with higher repetition rates (e.g, 100 kHz) can significantly reduce this acquisition time, opening the possibility of obtaining these structures as a function of pump-probe delay within a reasonable data acquisition time.\\

\clearpage
\section{\label{sec:epichlorohydrin} Additional results on epichlorohydrin}

\begin{figure}[h]
    \includegraphics[width=\columnwidth]{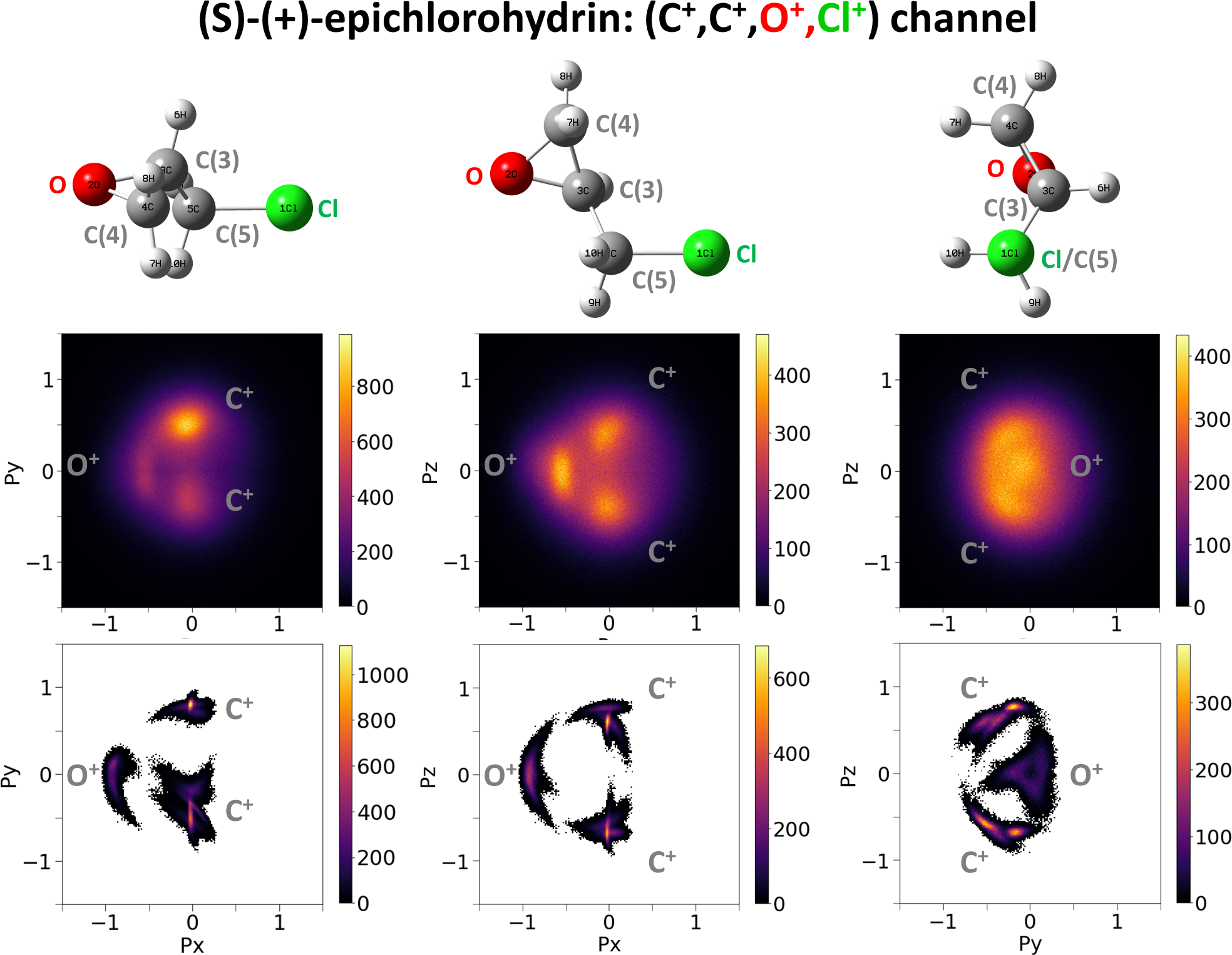}
    \caption{\textbf{Results for epichlorohydrin.} \textbf{Top:} Molecular views: The ball-and-stick model of epichlorohydrin in the equilibrium geometry is rotated such that the C(5)-Cl bond is along the $x$-axis, and that the line connecting C(5) and C(4) points in the positive $z$ direction and lies in the $xz$-plane. The $y$-axis is determined by the right-hand rule. \textbf{Middle:} Different projections of the measured momenta normalized to the momentum of $\mathrm{Cl^+}$. The momentum of $\mathrm{Cl^+}$ (not plotted) is along the $x$ axis. We used the momentum of the first $\mathrm{C^+}$ to define the $xy$-plane, since the momentum of $\mathrm{O^+}$ is almost back-to-back with the momentum of $\mathrm{Cl^+}$ as seen in the figure above where $\mathrm{O^+}$ has high magnitude of $p_x$ but its $p_y$ and $p_z$ distributions are centered around zero. The $xy-$reference ion (first $\mathrm{C^+}$) is plotted in the $p_xp_y$, but not $p_zp_x$ and $p_zp_y$ projections. Using the second $\mathrm{C^+}$ as the $xy$-reference yields almost identical results. \textbf{Bottom:} Simulations of the projections in the middle panels using a classical Coulomb explosion model. Since we do not distinguish the carbons in the experiments (the first detected carbon used as the $xy$-reference can be any of the three carbons in the molecule), we sum up all three simulated patterns where $\mathrm{Cl^+}$ defines the $x$-axis and $\mathrm{C(3)}$, $\mathrm{C(5)}$, $\mathrm{C(4)}$ alternatively defines the $xy$-plane.}
    \label{fig:epichlorohydrin_results}
\end{figure}

Figure~\ref{fig:epichlorohydrin_results} shows the ball-and-stick model and the measured and simulated CEI patterns for epichlorohydrin (the ring-chain molecule). The measured momentum images are obtained from the ($\mathrm{C^+}$, $\mathrm{C^+}$, $\mathrm{O^+}$, $\mathrm{Cl^+}$) 4-fold coincidence channel. The frame of reference is defined by the $\mathrm{Cl^+}$ and the first $\mathrm{C^+}$ momenta; the $\mathrm{O^+}$ and the other $\mathrm{C^+}$ ions are plotted in this frame.

\clearpage
\section{\label{sec 3-chloro-1-propanol} Additional results on 3-chloro-1-propanol}

\begin{figure}[h]
    \includegraphics[width=0.85\columnwidth]{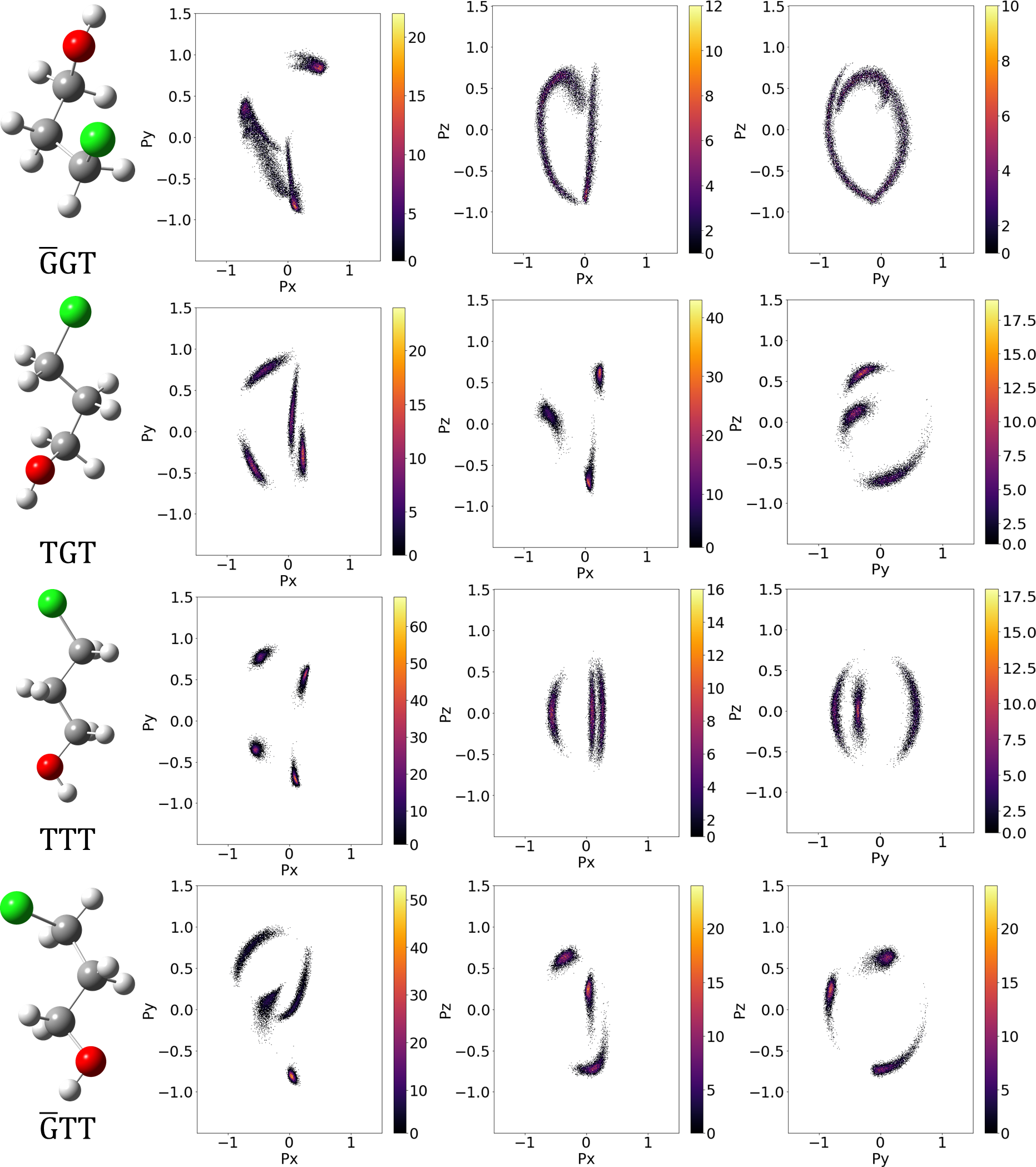}
    \caption{\textbf{Simulated momentum images of 3-chloro-1-propanol for different conformers.} Open-chain and ring-chain structures have single bonds that are able to rotate, leading to the existence of multiple conformers with low energy barriers \cite{Fuller1982, Richardson1997, Badawi2008, Wang2000, Endo2020}. Our simulations show that other conformers, with similar energies but different geometries, give very distinct momentum patterns different from the experimental results. In these momentum images, the momentum of $\mathrm{Cl^+}$ defines the $x$-axis, and the momentum of $\mathrm{O^+}$ defines the $xy$-plane. The naming convention for these conformers follows Ref.~\cite{Fuller1982, Richardson1997, Badawi2008}.}
    \label{fig:3_chloro_1_propanol_conformers}
\end{figure}

\begin{figure}[h]
    \includegraphics[width=0.8\columnwidth]{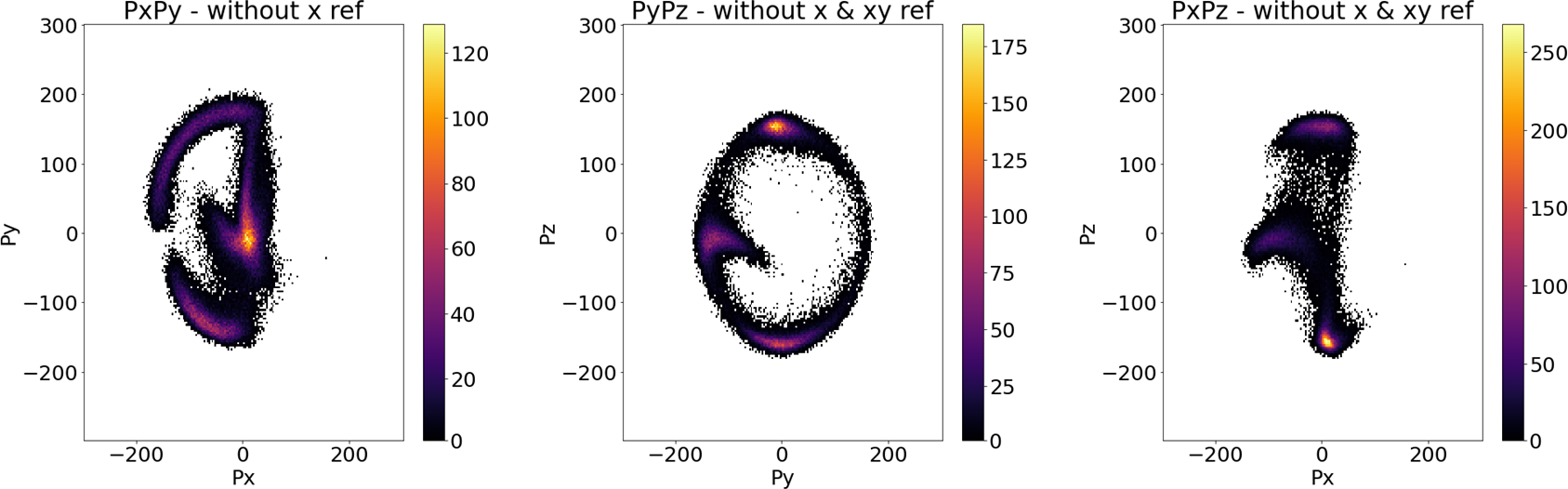}
    \caption{\textbf{Simulated momentum images of 3-chloro-1-propanol for an initial temperature of 3000~K ($\approx$~259~meV).} In these momentum images, the momentum of $\mathrm{Cl^+}$ defines the $x$-axis, and the momentum of $\mathrm{O^+}$ defines the $xy$-plane.}
    \label{fig:3_chloro_1_propanol_sim_3000K}
\end{figure}

\begin{figure}[h]
    \includegraphics[width=\columnwidth]{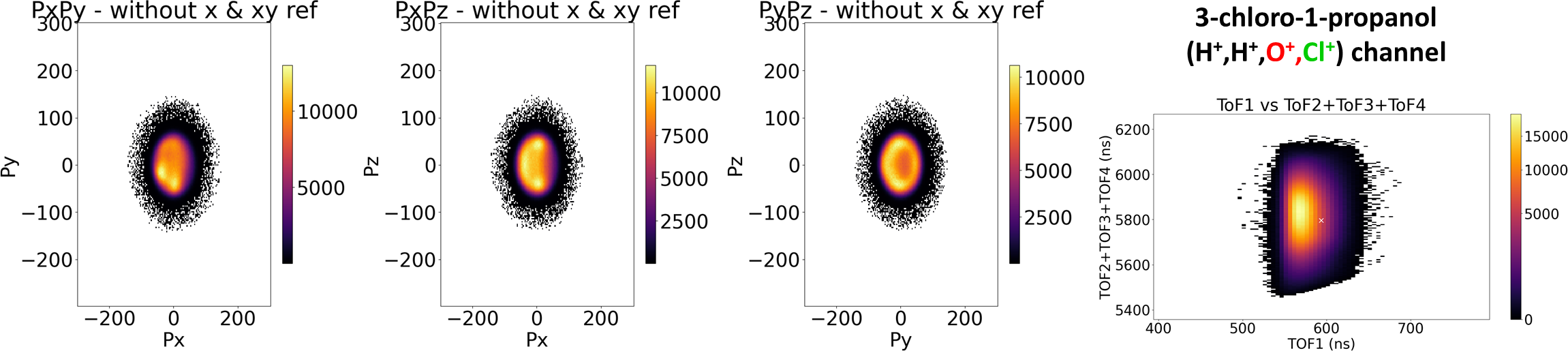}
    \caption{\textbf{Momentum images of protons from 3-chloro-1-propanol obtained from the 4-fold $\mathrm{(H^+,H^+,O^+,Cl^+)}$ coincidence channel.} In these momentum images, the momentum of $\mathrm{Cl^+}$ defines the $x$-axis, and the momentum of $\mathrm{O^+}$ defines the $xy$-plane.}
    \label{fig:3_chloro_1_propanol_protons}
\end{figure}

This section provides additional results on 3-chloro-1-propanol including simulated CEI patterns of other rotational conformers (Fig.~\ref{fig:3_chloro_1_propanol_conformers}), simulated CEI patterns of the lowest energy conformer at 3000 K (Fig.~\ref{fig:3_chloro_1_propanol_sim_3000K}), and the measured momentum images of protons (Fig.~\ref{fig:3_chloro_1_propanol_protons}). Figure~\ref{fig:3_chloro_1_propanol_conformers} shows that other conformers, with similar energies but different geometries, give very distinctive momentum patterns different from the experimental results. Figure~\ref{fig:3_chloro_1_propanol_sim_3000K} demonstrates the effect of vibrational excitation in broadening the momentum spectra. Figure~\ref{fig:3_chloro_1_propanol_protons} shows that we obtain some, but not clear, structures of protons under our experimental conditions.

Figure~\ref{fig:3_chloro_1_propanol_sim_3000K} shows that at a higher initial temperature, the momentum images start to show diffuse stripes connecting the $\mathrm{C^+}$ momenta similar to the experimental patterns presented in the main text. This initial distribution is not a thermal distribution but a broadening of the vibrational distributions. This observation suggests that these stripes, which are also seen in the experimental data, could come from vibrational excitation during the strong-field ionization process, spanning different geometries induced by different vibrational modes.

The momentum patterns of protons in 3-chloro-1-propanol are less clear as compared to isoxazole. We cannot assign the correspondences of individual hydrogens in the real space to the proton distributions in the momentum space. This observation suggests that the hydrogen atoms may undergo a more significant motion during the ionization and fragmentation processes. The presence of multiple hydrogens in various orientations in 3-chloro-1-propanol might also hinder clear visualization of the CEI patterns, given that the momentum distributions of protons from isoxazole as shown in the main text are narrow and well localized in the molecular plane ($xy-$plane) but exhibit significantly broader out-of-plane distributions (see $xz$ and $yz$ planes).

\clearpage

\section{\label{TOF} Time-of-flight and position versus time-of-flight spectra}

\begin{figure*}[h]
    \includegraphics[width=\columnwidth]{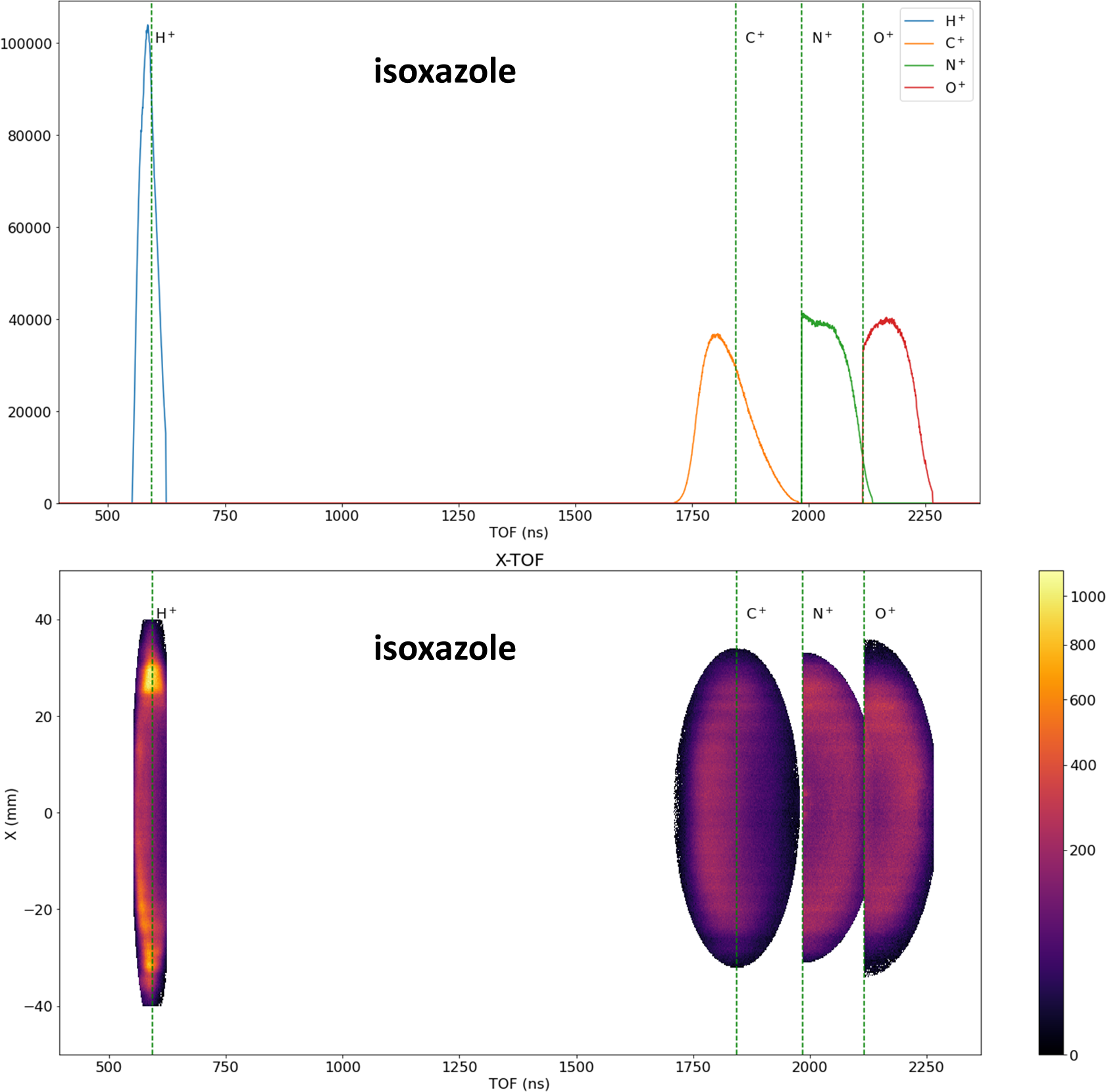}
    \caption{Time-of-flight and position versus time-of-flight spectra of isoxazole obtained from the 4-fold $\mathrm{(H^+,C^+,N^+,O^+)}$ coincidence channel.}
    \label{fig:tof_isoxazole}
\end{figure*}

\begin{figure*}[h]
    \includegraphics[width=0.6\columnwidth]{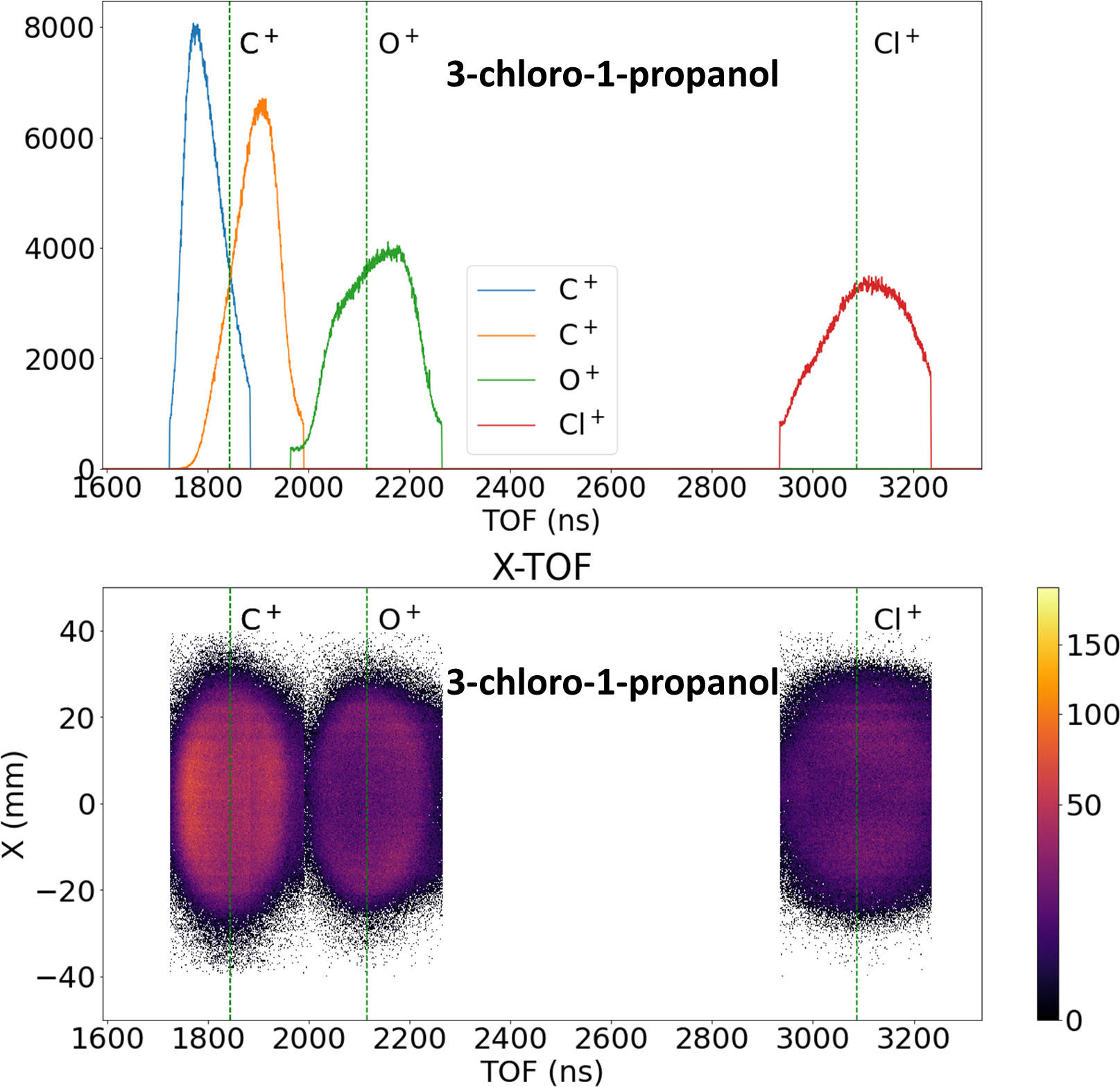}
    \caption{Time-of-flight and position versus time-of-flight spectra of 3-chloro-1-propanol obtained from the 4-fold $\mathrm{(C^+,C^+,O^+,Cl^+)}$ coincidence channel.}
    \label{fig:tof_3_chloro_1_propanol}
\end{figure*}

\begin{figure*}[h]
    \includegraphics[width=0.6\columnwidth]{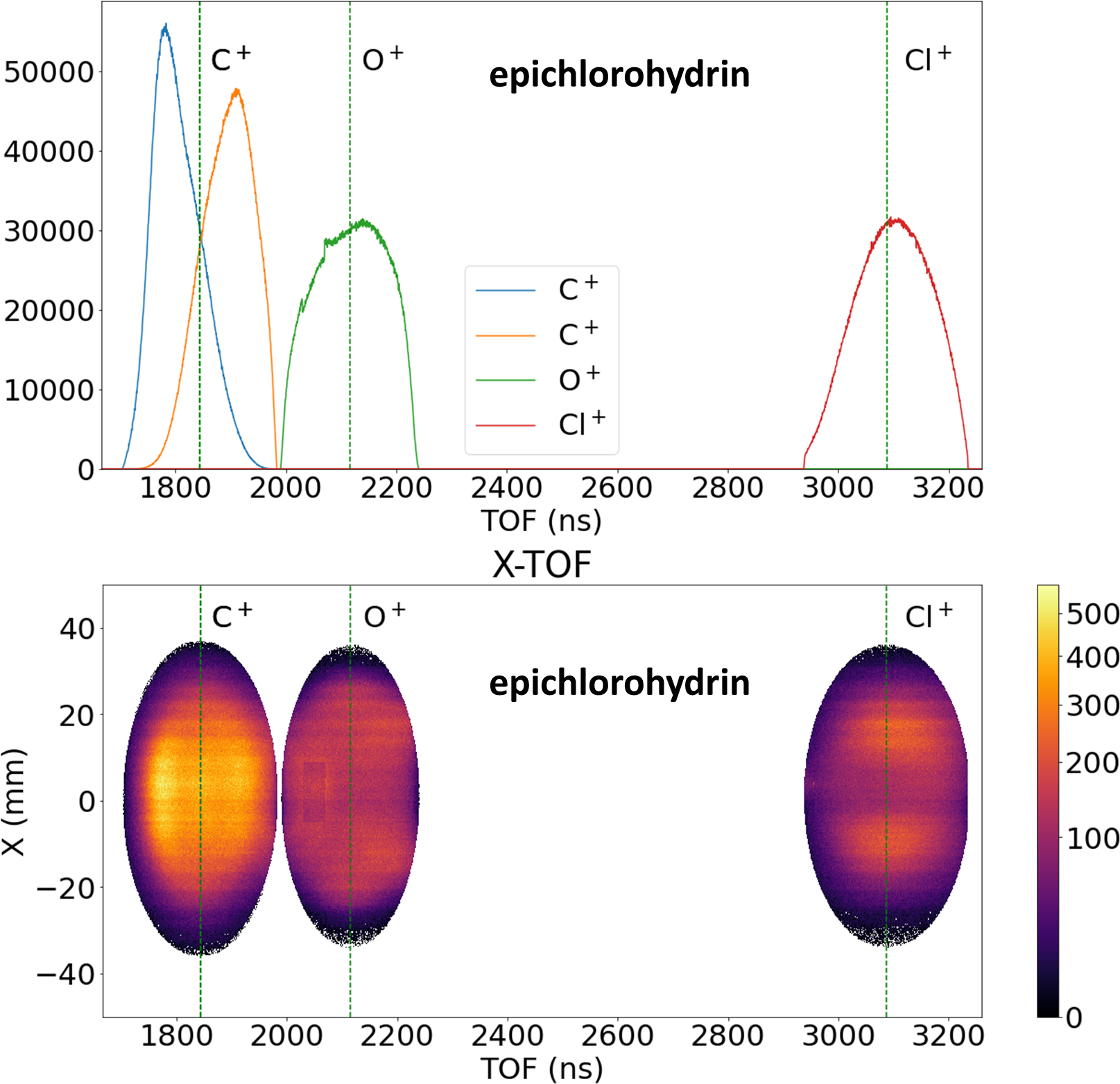}
    \caption{Time-of-flight and position versus time-of-flight spectra of epichlorohydrin obtained from the 4-fold $\mathrm{(C^+,C^+,O^+,Cl^+)}$ coincidence channel.}
    \label{fig:tof_ech}
\end{figure*}

This section provides the time-of-flight and position versus time-of-flight spectra of several coincidence channels presented in the main text for isoxazole (Fig.~\ref{fig:tof_isoxazole}), 3-chloro-1-propanol (Fig.~\ref{fig:tof_3_chloro_1_propanol}) and epichlorohydrin (Fig.~\ref{fig:tof_ech}). Hits from each species are gated based on their time-of-flight and also impact positions on the detector (to slightly reduce the diffused background or negate contributions from water or air). For isoxazole, we only use the right half time-of-flights of $\mathrm{N^+}$ and $\mathrm{O^+}$ to avoid the overlapping between $\mathrm{C^+}$, $\mathrm{N^+}$ and $\mathrm{O^+}$ ions (see Sec.~\ref{sec:coinc_channels} for a discussion on the effect of these gates). The “half-moon" gate is not required for 3-chloro-1-propanol and epichlorohydrin since the detected fragment ions are well separated on the time-of-flight axis.

\clearpage

\section{\label{GEOM} Equilibrium geometry of molecules used for Coulomb explosion simulations}

This section provides the equilibrium geometry of molecules used for Coulomb explosion simulations. The geometry of 3-chloro-1-propanol conformer is adapted from Ref.~\cite{Fuller1982, Richardson1997, Badawi2008}. The geometry of epichlorohydrin conformer is adapted from Ref.~\cite{Wang2000, Endo2020}. 

\begin{table}[h]
\caption{\label{tab:isoxazole} Equilibrium geometry of isoxazole optimized at the B3LYP/aug-cc-pvdz level.}
\begin{tabular}{|ccrrc|}
\hline
\multicolumn{5}{|c|}{isoxazole}                                                                                                              \\ \hline\hline
\multicolumn{1}{|l|}{}  & \multicolumn{1}{c|}{Atom} & \multicolumn{1}{c|}{x}        & \multicolumn{1}{c|}{y}        & \multicolumn{1}{c|}{z} \\ \hline
\multicolumn{1}{|c|}{1} & \multicolumn{1}{c|}{C}    & \multicolumn{1}{r|}{-3.38504} & \multicolumn{1}{r|}{1.85456}  & 0                      \\ \hline
\multicolumn{1}{|c|}{2} & \multicolumn{1}{c|}{C}    & \multicolumn{1}{r|}{-1.99880} & \multicolumn{1}{r|}{1.62072}  & 0                      \\ \hline
\multicolumn{1}{|c|}{3} & \multicolumn{1}{c|}{C}    & \multicolumn{1}{r|}{-1.89721} & \multicolumn{1}{r|}{0.25412}  & 0                      \\ \hline
\multicolumn{1}{|c|}{4} & \multicolumn{1}{c|}{H}    & \multicolumn{1}{r|}{-1.19389} & \multicolumn{1}{r|}{2.33914}  & 0                      \\ \hline
\multicolumn{1}{|c|}{5} & \multicolumn{1}{c|}{N}    & \multicolumn{1}{r|}{-4.08579} & \multicolumn{1}{r|}{0.71797}  & 0                      \\ \hline
\multicolumn{1}{|c|}{6} & \multicolumn{1}{c|}{H}    & \multicolumn{1}{r|}{-3.91694} & \multicolumn{1}{r|}{2.79554}  & 0                      \\ \hline
\multicolumn{1}{|c|}{7} & \multicolumn{1}{c|}{O}    & \multicolumn{1}{r|}{-3.13814} & \multicolumn{1}{r|}{-0.29474} & 0                      \\ \hline
\multicolumn{1}{|c|}{8} & \multicolumn{1}{c|}{H}    & \multicolumn{1}{r|}{-1.07633} & \multicolumn{1}{r|}{-0.44730} & 0                      \\ \hline
\end{tabular}
\end{table}

\begin{table}[h]
\caption{\label{tab:3-chloro-1-propanol} Equilibrium geometry of 3-chloro-1-propanol optimized at the B3LYP/aug-cc-pvdz level.}

\begin{tabular}{|ccrrr|}
\hline
\multicolumn{5}{|c|}{3-chloro-1-propanol}                                                                                                     \\ \hline
\multicolumn{1}{|l|}{}   & \multicolumn{1}{c|}{Atom} & \multicolumn{1}{c|}{x}        & \multicolumn{1}{c|}{y}        & \multicolumn{1}{c|}{z} \\ \hline
\multicolumn{1}{|c|}{1}  & \multicolumn{1}{c|}{Cl}   & \multicolumn{1}{r|}{1.81375}  & \multicolumn{1}{r|}{-0.44317} & -0.11624               \\ \hline
\multicolumn{1}{|c|}{2}  & \multicolumn{1}{c|}{O}    & \multicolumn{1}{r|}{-2.09120} & \multicolumn{1}{r|}{-0.56939} & 0.57725                \\ \hline
\multicolumn{1}{|c|}{3}  & \multicolumn{1}{c|}{C}    & \multicolumn{1}{r|}{-0.54867} & \multicolumn{1}{r|}{1.01743}  & -0.32061               \\ \hline
\multicolumn{1}{|c|}{4}  & \multicolumn{1}{c|}{C}    & \multicolumn{1}{r|}{0.58777}  & \multicolumn{1}{r|}{0.69071}  & 0.63709                \\ \hline
\multicolumn{1}{|c|}{5}  & \multicolumn{1}{c|}{C}    & \multicolumn{1}{r|}{-1.46011} & \multicolumn{1}{r|}{-0.15824} & -0.64254               \\ \hline
\multicolumn{1}{|c|}{6}  & \multicolumn{1}{c|}{H}    & \multicolumn{1}{r|}{-1.15328} & \multicolumn{1}{r|}{1.80802}  & 0.15313                \\ \hline
\multicolumn{1}{|c|}{7}  & \multicolumn{1}{c|}{H}    & \multicolumn{1}{r|}{-0.14451} & \multicolumn{1}{r|}{1.42965}  & -1.25576               \\ \hline
\multicolumn{1}{|c|}{8}  & \multicolumn{1}{c|}{H}    & \multicolumn{1}{r|}{-2.21666} & \multicolumn{1}{r|}{0.15805}  & -1.37985               \\ \hline
\multicolumn{1}{|c|}{9}  & \multicolumn{1}{c|}{H}    & \multicolumn{1}{r|}{-0.87530} & \multicolumn{1}{r|}{-0.98471} & -1.07563               \\ \hline
\multicolumn{1}{|c|}{10} & \multicolumn{1}{c|}{H}    & \multicolumn{1}{r|}{1.15158}  & \multicolumn{1}{r|}{1.58739}  & 0.91084                \\ \hline
\multicolumn{1}{|c|}{11} & \multicolumn{1}{c|}{H}    & \multicolumn{1}{r|}{0.22564}  & \multicolumn{1}{r|}{0.18531}  & 1.53653                \\ \hline
\multicolumn{1}{|c|}{12} & \multicolumn{1}{c|}{H}    & \multicolumn{1}{r|}{-2.56556} & \multicolumn{1}{r|}{-1.39408} & 0.42512                \\ \hline
\end{tabular}
\end{table}

\begin{table}[h]
\caption{\label{tab:epichlorohydrin} Equilibrium geometry of epichlorohydrin optimized at the B3LYP/aug-cc-pvdz level.}

\begin{tabular}{|ccrrr|}
\hline
\multicolumn{5}{|c|}{epichlorohydrin}                                                                                                         \\ \hline
\multicolumn{1}{|l|}{}   & \multicolumn{1}{c|}{Atom} & \multicolumn{1}{c|}{x}        & \multicolumn{1}{c|}{y}        & \multicolumn{1}{c|}{z} \\ \hline
\multicolumn{1}{|c|}{1}  & \multicolumn{1}{c|}{Cl}   & \multicolumn{1}{r|}{1.90603}  & \multicolumn{1}{r|}{0.22967}  & -0.03337               \\ \hline
\multicolumn{1}{|c|}{2}  & \multicolumn{1}{c|}{O}    & \multicolumn{1}{r|}{-2.06694} & \multicolumn{1}{r|}{-0.52063} & 0.00520                \\ \hline
\multicolumn{1}{|c|}{3}  & \multicolumn{1}{c|}{C}    & \multicolumn{1}{r|}{-0.77103} & \multicolumn{1}{r|}{-0.10592} & -0.46384               \\ \hline
\multicolumn{1}{|c|}{4}  & \multicolumn{1}{c|}{C}    & \multicolumn{1}{r|}{-1.66573} & \multicolumn{1}{r|}{0.85322}  & 0.19592                \\ \hline
\multicolumn{1}{|c|}{5}  & \multicolumn{1}{c|}{C}    & \multicolumn{1}{r|}{0.38643}  & \multicolumn{1}{r|}{-0.73097} & 0.26711                \\ \hline
\multicolumn{1}{|c|}{6}  & \multicolumn{1}{c|}{H}    & \multicolumn{1}{r|}{-0.67399} & \multicolumn{1}{r|}{-0.06702} & -1.55264               \\ \hline
\multicolumn{1}{|c|}{7}  & \multicolumn{1}{c|}{H}    & \multicolumn{1}{r|}{-1.47679} & \multicolumn{1}{r|}{1.13817}  & 1.23410                \\ \hline
\multicolumn{1}{|c|}{8}  & \multicolumn{1}{c|}{H}    & \multicolumn{1}{r|}{-2.21539} & \multicolumn{1}{r|}{1.57929}  & -0.40708               \\ \hline
\multicolumn{1}{|c|}{9}  & \multicolumn{1}{c|}{H}    & \multicolumn{1}{r|}{0.58739}  & \multicolumn{1}{r|}{-1.74602} & -0.09132               \\ \hline
\multicolumn{1}{|c|}{10} & \multicolumn{1}{c|}{H}    & \multicolumn{1}{r|}{0.21379}  & \multicolumn{1}{r|}{-0.74172} & 1.34755                \\ \hline
\end{tabular}
\end{table}

\clearpage
\bibliography{CEI_single_pulse_2022}